\documentclass[twocolumn,resetfootnote]{aastex701}

\usepackage{amsmath}
\usepackage{booktabs}
\usepackage{multirow}
\usepackage[table]{xcolor}
\usepackage{mathtools}
\usepackage[T1]{fontenc}

\begin{document}
\title{Substructure evolution from protoplanetary to debris disks driven by mutually gravitating planetesimals and implications on Kepler resonances and free-floating planets}

\author[orcid=0000-0002-2106-0403,sname=Han]{Yinuo Han}
\affiliation{Division of Geological and Planetary Sciences, California Institute of Technology, 1200 E. California Blvd., Pasadena, CA 91125, USA}
\email[show]{yinuo@caltech.edu}

\author[orcid=0000-0002-7094-7908,sname=Batygin]{Konstantin Batygin}
\affiliation{Division of Geological and Planetary Sciences, California Institute of Technology, 1200 E. California Blvd., Pasadena, CA 91125, USA}
\email{k.batygin@gmail.com}

\author[orcid=0000-0002-8958-0683,sname=Dai]{Fei Dai}
\affiliation{Institute for Astronomy, University of Hawai'i, 2680 Woodlawn Drive, Honolulu, HI 96822, USA}
\email{fdai@hawaii.edu}

\begin{abstract}
Motivated by recent observations suggesting rings found in debris disks are wider than those in protoplanetary disks, we consider a picture in which planetesimals formed in radially narrow dust traps radially diffuse into debris rings via mutual scattering. Under this picture, evolving the fractional widths ($\Delta r/r$) of resolved debris rings back to a few Myr according to the theoretical $t^{1/5}$ evolutionary trajectory reproduces the protoplanetary ring distribution. We inferred the product of the ring mass and individual planetesimal mass required to reach the observed debris ring widths at their ages, finding that $M_\mathrm{disk}\times m$ ranges from $10^{-3}$ to $10^{3}\,M_\oplus^2$. The distribution of $M_\mathrm{disk} \times m$ appears to correlate with the stellar mass, peaking at 1.5 to 2\,$M_\odot$, which resembles the stellar mass dependence of the giant exoplanet occurrence rate. The population of resolved debris rings lie close to the $\Delta r / r = 10\,h$ equipartition relation expected of a planetesimal ring that formed narrow, with typical resolved debris disks still expected to be broadening radially and vertically at present. If sufficiently massive ($\sim$10\,$M_\oplus$), this radial broadening can send a few Mercurys to the terrestrial region within 10\,Myr, making debris disks a plausible source of planetesimals disrupting resonant chains among Kepler planets. Within Gyr timescales, outer planetesimal belts can also eject $\sim$1\% of their mass into interstellar space if they consist of Moon-sized bodies or above, suggesting that the slow and steady intrinsic evolution of massive debris disks could contribute to the interstellar free-floating population of terrestrial-planet-sized bodies.
\end{abstract}

\keywords{\uat{Circumstellar disks}{235} --- \uat{Protoplanetary disks}{1300} --- \uat{Debris disks}{363} --- \uat{Planetary dynamics}{2173} --- \uat{Orbits}{1184} --- \uat{N-body simulations}{1083}}


\section{Introduction} \label{sec:introduction}
The gas- and dust-rich environment within protoplanetary disks is understood to be the central theatre of planet formation \citep{Williams2011, Armitage2011}. From these primordial disks assemble systems of planets and planetesimal belts, the latter of which are referred to as debris disks upon dispersal of primordial material \citep{Wyatt2008, Hughes2018}. 

With the capabilities of the Atacama Large Millimeter/submillimeter Array (ALMA) over the past decade, a substantial volume of observations have resolved the detailed structure of both protoplanetary \citep{Andrews2018, Bae2023} and debris disks \citep{Matra2025, Marino2026} at spatial resolutions down to a few au. These observations have revealed that while ring-like structures are common at both disk stages, they appear to be structurally distinct. In particular, debris disks appear to be ``fractionally broader'' than rings in protoplanetary disks as measured by $\Delta r/r$ \citep{Matra2025}, where $r$ is the radius from the star and $\Delta r$ the full width at half maximum (FWHM) of the ring's radial profile. Recent high-resolution ALMA observations \citep{Marino2026} have shown that many individual rings in debris disks are narrower than previously thought, however a notable fraction ($\sim$40\%) is still thought to be significantly broader than those in protoplanetary disks \citep{Han2026}.

Such a discrepancy would appear to require an explanation. Protoplanetary rings have long been hypothesised to be a key site of planetesimal formation, where millimetre-sized dust grains gathered by pressure bumps \citep{Stadler2025, Zhao2025} collapse into planetesimals with the help of two-fluid instabilities \citep{Tatarelli2026}. 
If these rings are the main site of planetesimal formation \citep{Jiang2023}, the planetesimal belts formed are expected to inherit the spatial distribution of narrow protoplanetary dust rings. The fact that a significant proportion of debris disks observed are broader would then suggest a prevalent mechanism that widens these rings at formation or subsequently over time. 

Motivated by these observations, we seek a simple mechanism that could provide an (at least partial) explanation for such a radial evolution. 
Previous studies have suggested explanations such as pertuerbations from planets \citep{Gomes2005, Matra2025}, or protoplanetary dust traps that migrate under the influence of protoplanets \citep{Miller2021}. Here we explore a simpler solution that does not require external perturbers, aiming instead to characterise the intrinsic radial evolution of a planetesimal belt to leading order following its formation in a narrow ring, and to compare theoretical expectations with resolved imaging.

In the following sections, we begin by considering analytical scaling relations of mutually gravitating disk particles in a central potential in Sec.~\ref{sec:scaling}, which are compared to N-body simulations in Sec.~\ref{sec:n-body}. We then apply these theoretical relations to observations in Sec.~\ref{sec:observations} and discuss their wider implications in Sec.~\ref{sec:discussion}, including on the protoplanetary to debris disk evolution, the disruption of orbital resonances observed among terrestrial planets, and the ejection of planet-sized bodies into interstellar space. These findings are summarised Sec.~\ref{sec:conclusions}.

\section{Scaling relations} \label{sec:scaling}
The aim of this section is to quantify to leading order the scaling relations between the radial width of an initially narrow planetesimal ring and other basic properties of the system, including primarily its age ($t$). We do so by considering the orbital elements of equal-mass ($m$) planetesimals in a ring, including their semimajor axis ($a$), eccentricity ($e$) and inclination ($i$), with each pair of planetesimals feeling each other's gravity. 

The dispersion of the three orbital elements are set by the dispersion of random velocities, $v$, defined relative to co-located circular orbits. Instead of dealing directly with $a$, we define the dimensionless quantity,
\begin{equation}
    \tilde{a} = \frac{\Delta a}{a_\mu} = \frac{a - a_\mu}{a_\mu},
\end{equation}
to characterize the departure of $a$ for a given planetesimal from the mean semimajor axis of all planetesimals in the ring, $a_\mu$. We can then write the three orbital elements of interest as 
\begin{equation}
    \label{eq:k}
    \begin{bmatrix}
    \tilde{a}\\
    e\\
    i
    \end{bmatrix}
     = 
    \begin{bmatrix}
    k_{\tilde{a}}\\
    k_e\\
    k_i
    \end{bmatrix}
    \frac{v}{v_\mathrm{K}},
\end{equation}
where $v_\mathrm{K}$ is the circular Keplerian velocity of the planetesimal ring at $a_\mu$ and $k$'s are constants of order unity. 

A range of physical effects could influence the evolution of $v$, however here we aim to explore the leading-order behaviour of $v$ within a single-population planetesimal disk via the dominant effect of viscous stirring \citep{Stewart1988, Goldreich2004}. In realistic disks, heating or cooling via dynamical friction (in disks with multiple planetesimal populations, \citealp{Ida1990}) and damping via inelastic collisions could influence $v$. While we do not model these higher-order effects, we account for the upper limit of the velocity dispersion reachable via viscous stirring, beyond which physical collisions become dominant and random velocities cease to increase. We neglect gas drag as we primarily apply our analysis to planetesimal belts after primordial gas disk dispersal \citep{Pearce2025, MacManamon2026}. 

The viscous stirring rate is summarised with order-of-magnitude arguments in \citet{Goldreich2004} with an $n \sigma v$ approach, where $n$ is the number density of planetesimals within a 3D volume, $\sigma$ is the cross-sectional area of interaction, and $v$ is the relative velocity between planetesimals. 

In the regime where the velocity dispersion lies between the Hill velocity and escape velocity, i.e., $v_H < v < v_\mathrm{esc}$, the primary outcome of close encounters between planetesimals is gravitational deflections that excite $v$ \citep{Goldreich2004}. In the outer planetary system, $v_H < v$ is satisfied for Pluto-sized bodies as long as $e \gtrsim 10^{-3}$. Assuming a Safronov number of $\sim (v_\mathrm{esc}/v)^2 \gg 1$, the doubling rate of $v$ then scales as
\begin{equation}
    \label{eq:vdvdt}
    \frac{1}{v} \frac{dv}{dt} \propto n \left[ \pi R^2 \left( \frac{v_\mathrm{esc}}{v} \right)^4 \right] v,
\end{equation}
where $R$ is the planetesimal radius, $v_\mathrm{esc}$ is the planetesimal surface escape velocity and $t$ is the age of the system. 

This is effectively the two-body relaxation rate, for which $\left( v / \Delta v \right)^2$ randomly oriented kicks, which occur at a rate of $n (\pi b^2) v$, is required for kicks of magnitude $\Delta v = t_\mathrm{cross}(Gm/b^2)$ to multiply $v$ by order unity, where $b$ is the impact parameter and $t_\mathrm{cross} = b/v$ is the crossing timescale. 
The $\left( v_\mathrm{esc} / v \right)^2$ enhancement in the critical impact parameter for momentum exchange is larger than the critical impact parameter required for an impact with gravitational focussing. This critical radius is also the separation between two planetesimals where their gravitational potential energy is equal in magnitude to the random kinetic energy. 

Assuming that the scale height $h \propto v/v_\mathrm{K}$, Eq.~\eqref{eq:vdvdt} can equivalently be written as
\begin{equation}
    \label{eq:vdvdt2}
    \frac{1}{v} \frac{dv}{dt} \propto \frac{\Sigma \Omega_\mathrm{K}}{\rho R} \left( \frac{v_\mathrm{esc}}{v} \right)^4,
\end{equation}
where $R$ and $\rho$ are the radius and mass density of an individual planetesimal, $v_\mathrm{esc}$ is the planetesimal surface escape velocity, $\Sigma$ is the surface density of the disk and $\Omega_\mathrm{K}$ is the Keplerian angular velocity. 

This is a particularly useful form when dealing with a configuration consisting of smaller bodies evolving under the influence of larger stirrers, in which $\Sigma$ is dominated by massive stirrers whose spatial distribution remains largely constant over time. Integrating Eq.~\eqref{eq:vdvdt2}, such a setup results in the velocity dispersion of the small bodies to scale as $v \propto t^{1/4}$, as has been extensively explored and applied in the literature \citep{Stewart2000, Goldreich2004, Armitage2010, Kokubo2012, Pearce2025, Chiang2026}. 

However, here we are primarily interested in the velocity evolution of massive planetesimals themselves through mutual scattering, thus $\Sigma$ and $n$ depend on $v$. To account for this dependence, the number density of an initially narrow ring can be approximated as
\begin{equation}
    \label{eq:n}
    n \propto \frac{N}{ (2 \pi a_\mu) \left [ a_\mu \left( \tilde{a}^2 + e^2 \right)^{1/2} \right ] (a_\mu i)},
\end{equation}
where $N$ is the number of planetesimals in the disk. 

Eq.~\eqref{eq:vdvdt} can then be written as
\begin{equation}
    \frac{1}{v} \frac{dv}{dt} \propto \frac{1}{k_i \left(k_{\tilde{a}}^2 + k_e^2 \right)^{1/2} } \frac{N}{a_\mu^3} R^2 v_\mathrm{K}^2 v_\mathrm{esc}^4 \frac{1}{v^5},
\end{equation}
Ignoring constants of order unity, $v$ thus evolves as
\begin{equation}
    \left( \frac{v}{v_\mathrm{K}} \right)^5 - \left( \frac{v_0}{v_\mathrm{K}} \right)^5 \propto N m^2 M_*^{-3/2} a_\mu^{-3/2} t,
\end{equation}
where $M_*$ is the stellar mass and $v_0$ is the random velocity dispersion at $t = 0$.   

The three orbital elements of interest are therefore expected to evolve as 
\begin{align}
    \label{eq:a}
    \tilde{a}^5(t) - \tilde{a}_0^5 &= \left[ B_{\tilde{a}} \xi(t) \right]^5,\\
    \label{eq:e}
    e^5(t) - e_0^5 &= \left[ B_e \xi(t) \right]^5,\\
    \label{eq:i}
    i^5(t) - i_0^5 &= \left[ B_i \xi(t) \right]^5,
\end{align}
where
\begin{equation}
    \label{eq:xi}
    \xi(t) = N^{1/5} m^{2/5} M_*^{-3/10} a_\mu^{-3/10} t^{1/5},
\end{equation}
and the $B$'s are constants of proportionality. 

Over sufficiently long timescales such that $v \gg v_0$, we thus expect $v \propto t^{1/5}$, which doubles more slowly than $v \propto t^{1/4}$ for any smaller bodies stirred by these massive bodies within the disk. This twice-longer doubling time further implies that the small bodies respond quickly, reaching equilibrium almost immediately, and are effectively dictated by the slow viscous evolution of the large bodies reflected in $\Sigma$ in Eq.~\eqref{eq:vdvdt2}. 

We will explore the implications of this intrinsic ``viscous spreading'' in more detail in subsequent sections, but we will first use N-body simulations to verify this simple analytic scaling and measure realistic constants of proportionality in Sec.~\ref{sec:n-body}.

\section{N-body simulations} \label{sec:n-body}
We performed N-body simulations to test the analytic scaling in Eq.~\eqref{eq:a}--\eqref{eq:xi}, tracking the time evolution of $\tilde{a}$, $e$ and $i$ of an initially narrow ring of equal-mass planetesimals. The general approach was to perform a baseline simulation (Run i) to establish the time dependence of the orbital elements, before performing additional simulations that tweak individual parameters compared to Run i to test the $N$, $m$, $M_\star$ and $a_\mu$ dependencies. These simulations are summarised in Table~\ref{tab:nbody}. Note that not all simulations may represent realistic or observationally known configurations (e.g., an M-type star with a planetesimal belt as far out as at 100\,au), but we exaggerate certain quantities to obtain more robust tests on scaling relations given the relatively shallow dependencies expected. 

As we will encounter various statistical distributions relating to variables such as $a$, $e$ and $i$, we explicitly denote the mean of a variable $x$ as $x_\mathrm{\mu}$, the standard deviation as $x_\mathrm{sd}$, the full width at half maximum as $x_\mathrm{FWHM}$ and the root-mean-square as $x_\mathrm{rms}$. For Rayleigh distributions, we denote the Rayleigh scale parameter as $x_\mathrm{\sigma}$ such that $x_\mathrm{\mu} = \sqrt{\pi/2} \, x_\mathrm{\sigma}$, $x_\mathrm{sd} = \sqrt{2 -\pi/2} \, x_\mathrm{\sigma}$ and $x_\mathrm{rms} = \sqrt{2} \,  x_\mathrm{\sigma}$. For normal distributions, the Gaussian scale parameter is equal to the standard deviation, i.e., $x_\mathrm{sd} = x_\sigma$.

\begin{table*}
\centering
\caption{N-body simulation parameters.}
\label{tab:nbody}
\begin{tabular}{lccccccccccc}
\toprule
Run & $N$ & $R$ & $m$ & $M_\mathrm{disk}$ & $a_\mu$ & $M_*$ & $\tilde{a}_0$ & $e_0$ & $i_0$ & $\Delta t$ & $t$ \\ 
 & & [km] & [$M_\oplus$] & [$M_\oplus$] & [au] & [$M_\odot$] & & & & [yr] & [Myr] \\ 
\midrule
i   & 1024 & 1000 & $1.9 \times 10^{-3}$ & 1.94 & 100 & 1   & 
\multirow{5}{2.1cm}{\centering Gaussian $\tilde{a}_{\sigma, 0} = 0.01 \sqrt{\pi}$} & 
\multirow{5}{1.7cm}{\centering Rayleigh $e_{\sigma, 0} = 0.01$} & 
\multirow{5}{1.7cm}{\centering Rayleigh $i_{\sigma, 0} = 0.005$} & 
33 & 150 \\ 
ii  & 512  & 1260 & $3.8 \times 10^{-3}$ & 1.94 & 100 & 1   & & & & 33 & 10 \\ 
iii & 512  & 1000 & $1.9 \times 10^{-3}$ & 0.97 & 100 & 1   & & & & 33 & 10 \\ 
iv  & 1024  & 1000 & $1.9 \times 10^{-3}$ & 1.94 & 50  & 1   & & & & 11 & 10 \\ 
v   & 1024  & 1000 & $1.9 \times 10^{-3}$ & 1.94 & 100 & 0.5 & & & & 33 & 10 \\ 
\bottomrule
\end{tabular}
\end{table*}

\subsection{Initial setup}
We focused on the case of equipartition when initialising a narrow ring of planetesimals. We will show subsequently in Sec.~\ref{sec:observations} that observed planetesimal belts (and indeed dust rings in protoplanetary rings) appear to be centred not too far off from equipartition. 

To determine the distribution and relative dispersion of $\tilde{a}$, $e$ and $i$ at equipartition, we performed a test run (Run T1 in Appendix~\ref{sec:equipartition}), in which $\tilde{a}$, $e$ and $i$ were drawn from identical uniform boxes between 0 and 0.01. The longitude of ascending node ($\Omega$), argument of periastron ($\omega$) and mean anomaly ($M$) were drawn uniformly between 0 and $2 \pi$, as is the case for all subsequent runs. We found that for a setup otherwise identical to Run~i, equipartition is rapidly established within $<0.1$\,Myr, beyond which $\tilde{a}$ remains normally distributed, $e$ and $i$ remain Rayleigh distributed, $\tilde{a}_\sigma$ = 1.4 $e_\mu$ and $e_\mu$ = 2 $i_\mu$. We therefore adopted these distributions and relative dispersions as the initial conditions for all runs in Table~\ref{tab:nbody}. 

To evolve the planetesimal ring, we used the N-body code GENGA (Gravitational ENcounters with Gpu Acceleration, \citealp{Grimm2014, Grimm2022}). 
GENGA uses a direct N-body integrator with the Bulirsch–Stoer method for bodies undergoing a close encounter, or a mixed variable symplectic integrator \citep{Wisdom1991} that allows for large time steps ($\sim1/30$ of an orbit) when bodies are far. We used the default critical radius for close encounters, defined as either 3 Hill radii or 0.4 times the distance traversed per time step, whichever is larger. 
Any collisions are assumed to be completely inelastic, although in practice the collision rate is low in our simulations (at most a few for 1000 bodies over 100\,Myr). We assumed the density of planetesimals to be $\rho = 2.7$\,g\,cm$^{-3}$ when defining planetesimals by their radius ($R$ in Table~\ref{tab:nbody}). 

\subsection{Evolution of a, e and i}
\label{sec:aei}

\begin{figure*}
    \centering
    \includegraphics[width=1.0\linewidth]{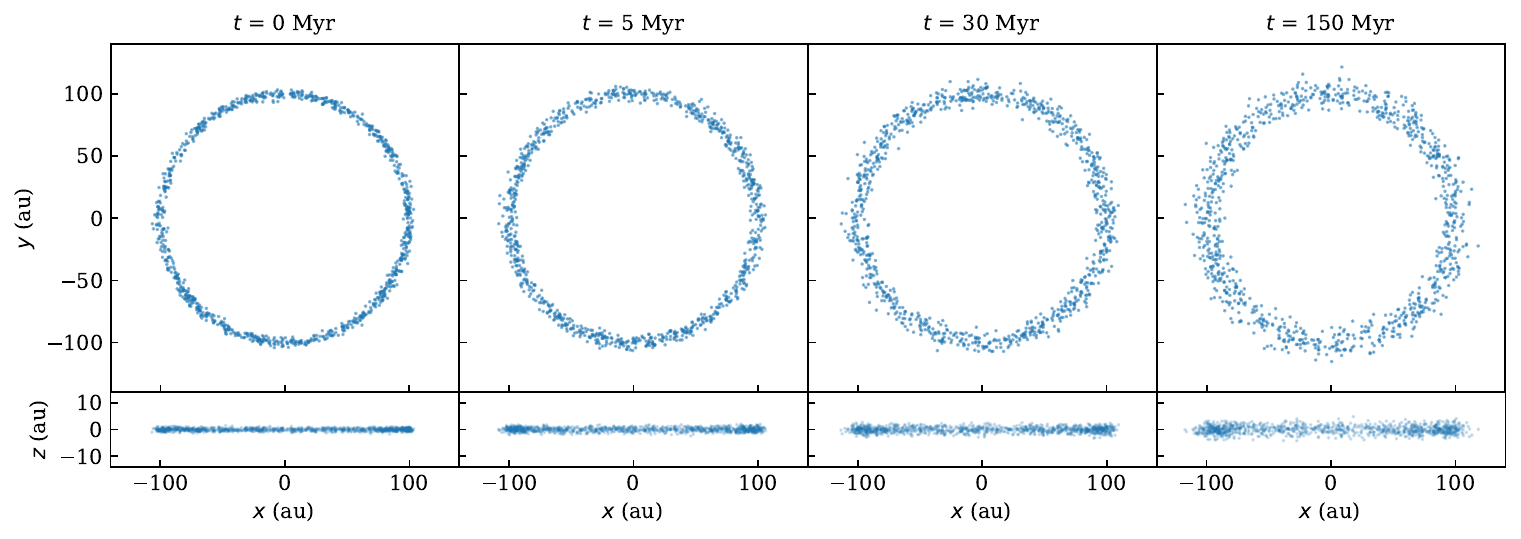}
    \caption{Snapshots from Run i showing the evolution of a narrow belt of planetesimals undergoing mutual gravitational deflections from close encounters.}
    \label{fig:snapshots}
\end{figure*}

\begin{figure*}
    \centering
    \includegraphics[width=1.0\linewidth]{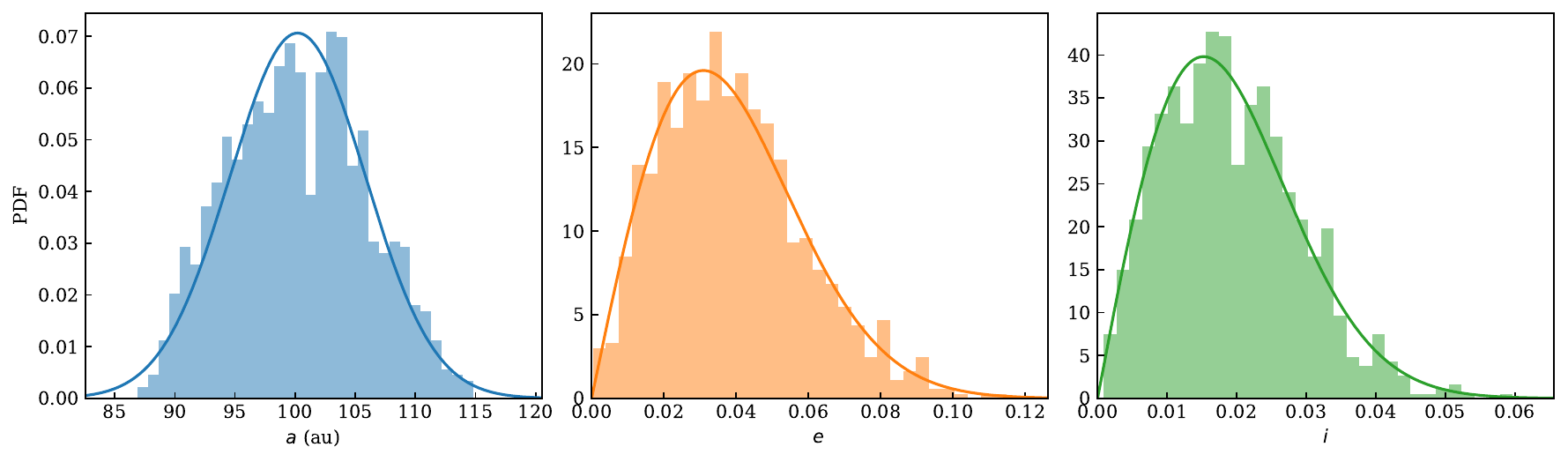}
    \caption{The probability density functions (PDFs) of $a$, $e$ and $i$ at 150\,Myr in Run i. The solid curves show the best-fit normal distribution for $a$ and the best-fit Rayleigh distributions for $e$ and $i$.}
    \label{fig:distributions}
\end{figure*}

We evolved the baseline case, Run i, to 150 Myr to track the evolution of a narrow belt of $10^3$ Pluto-sized bodies at 100\,au around a solar-mass star. Snapshots of the disk are displayed in Fig.~\ref{fig:snapshots}, which show the disk to spread radially and vertically as qualitatively expected. 

A snapshot of the underlying $\tilde{a}$, $e$ and $i$ distribution is shown in Fig.~\ref{fig:distributions}. The distribution of $e$ and $i$ are well-described as Rayleigh, as expected from a 2D random walk in ($e \cos{\omega}$, $e \sin{\omega}$) due to random kicks in 3D space for $e$, in which each component is normally distributed, and in ($i \cos{\Omega}$, $i \sin{\Omega}$) for $i$, where $\omega$ is the argument of periastron and $\Omega$ is the longitude of ascending node. This is a well-studied result in the literature \citep{Greenzweig1992, Ida1992}. In contrast, the distribution of $\tilde{a}$ is Gaussian. This appears to be less frequently the focus of previous work (but see high-resolution N-body simulations by \citealt{Woo2023}), but is expected from a 1D random walk in $a$ itself.

\begin{figure*}
    \centering
    \includegraphics[width=1.0\linewidth]{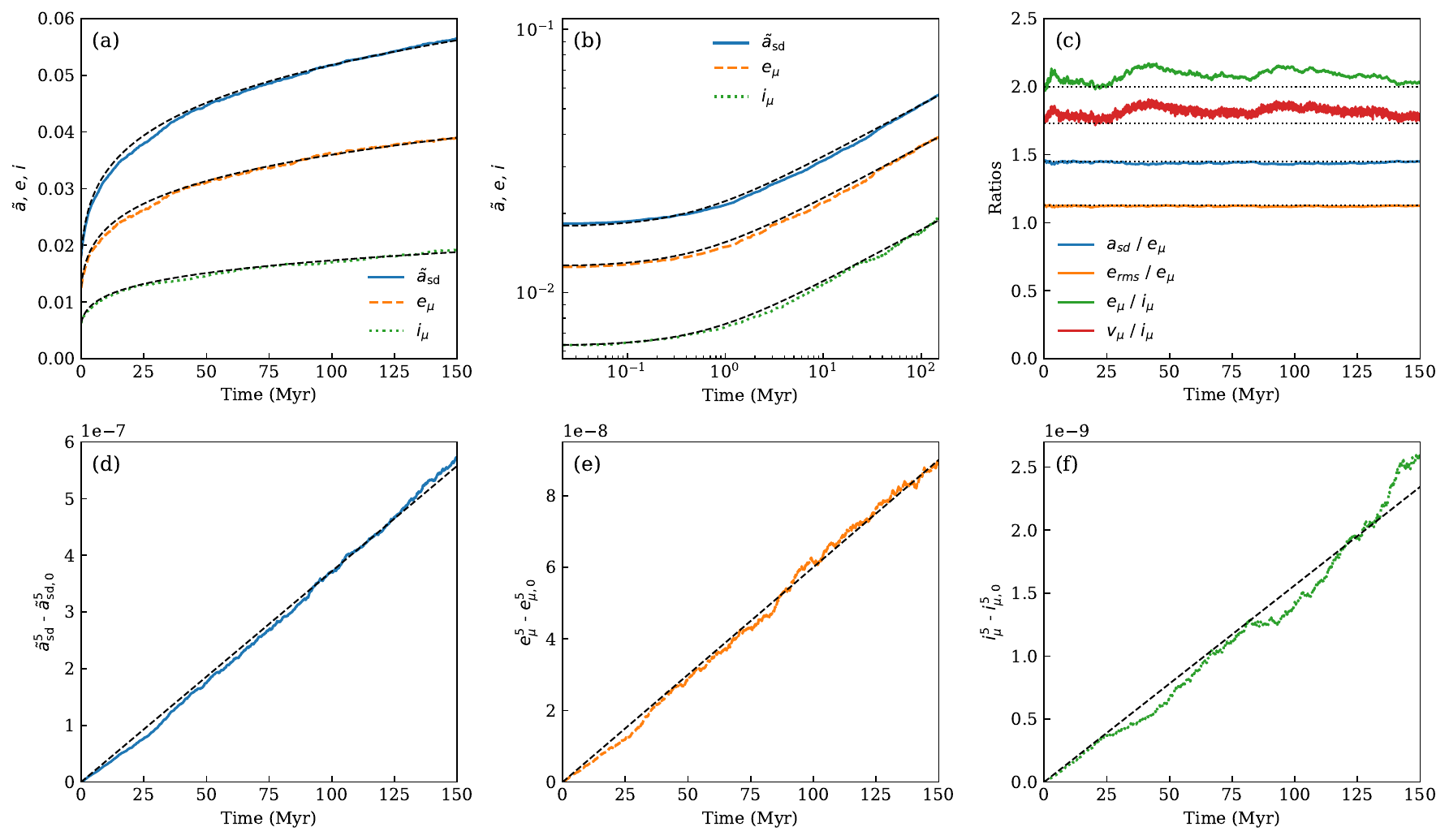}
    \caption{The evolution of $\tilde{a}$, $e$ and $i$ in Run i. Panels (a) and (b) show the three quantities on linear and logarithmic scales. Panel (c) shows ratios between $\tilde{a}$, $e$ and $i$ throughout the simulation. Panels (d)--(f) show $\tilde{a}$, $e$ and $i$ transformed according to Eq.~\eqref{eq:a}--\eqref{eq:i}, which are expected to evolve linearly with time. The dashed lines correspond to a $t^{1/5}$ model based on Eq.~\eqref{eq:a}--\eqref{eq:xi} with constants of proportionality ($A_{\tilde{a}}$, $A_e$ and $A_i$) fitted to the simulation. }
    \label{fig:aei}
\end{figure*}

To characterise the time evolution of the spread of $\tilde{a}$, $e$ and $i$, we use the standard deviation (sd) for $\tilde{a}$ given it is normally distributed, and the mean ($\mu$) for $e$ and $i$ since they are Rayleigh distributed. The evolution of the $\tilde{a}$, $e$ and $i$ dispersions are shown in Fig.~\ref{fig:aei}. 

We find that the dispersion of the three orbital elements evolve in synchrony (Fig.~\ref{fig:aei} panels a and b) as predicted in Eq.~\eqref{eq:a}--\eqref{eq:i}, with all three dictated by the excitation of random velocities. Throughout the simulation, the three quantities maintain constant ratios (panel c). The characteristic Rayleigh $e_\mathrm{rms}$ to $e_\mu$ ratio of $2/\sqrt{\pi}$ is maintained throughout the simulation, reflecting a stable Rayleigh distribution. The $e$ to $i$ ratio is on average 2.08 throughout the simulation, which is similar to the theoretical prediction of $e = 2i$ \citep{Ida1992, Armitage2010}. The mean $\tilde{a}_\sigma$ to $e_\mu$ ratio is 1.44. This is expected from energy and angular momentum conservation, where $E$ and $L$ given by
\begin{align}
    &E_i \propto - \frac{1}{a_0 (1 + \tilde{a}_i)},\\
    &L_i \propto \sqrt{a_0 (1 + \tilde{a}_i) (1 - e_i^2)} \, \cos{i_i},
\end{align}
expanded to second order in $\tilde{a}_i$, $e_i$ and $i_i$ requires 
\begin{equation}
    \sum_i \tilde{a}_i^2 = \frac{4}{3} \sum_i \left( e_i^2 + i_i^2 \right),
\end{equation}
or equivalently, $\tilde{a}_{\sigma} = 1.45 e_\mu$, given the $e$ to $i$ ratio that we find. 
Taken together, the equipartition distribution follows 
\begin{equation}
    \label{eq:aei}
    \tilde{a}_\sigma : e_\mu : i_\mu \approx 3:2:1. 
\end{equation}

Fig.~\ref{fig:aei}c also displays the normalised random velocity, $\tilde{v} = v / v_{\mathrm{K}}$, as a multiple of the inclination.  
The mean $\tilde{v}_\mu$ to $i_\mu$ ratio is 1.81, close to $\sqrt{3}$ as expected from Eq.~\ref{eq:aei} and $v = \sqrt{5/8 \, e^2 + 1/2 \, i^2}$ (Eq.~16 in \citealt{Lissauer1993}). The random velocity relative to a swarm of other planetesimals, $\tilde{v}_\mathrm{swm}$, is greater than $v$ by a factor of $\sqrt{2}$ (Eq.~17 in \citealt{Lissauer1993}).

To characterise the time evolution of the three orbital elements, we plotted the transformed quantities $\tilde{a}^5_{\sigma}(t) - \tilde{a}^5_{\sigma, 0}$, $e^5_{\mu}(t) - e^5_{\mu, 0}$ and $i^5_{\mu}(t) - i^5_{\mu, 0}$ in Fig.~\ref{fig:aei}e--f, which are expected to scale with $t$ from Eq.~\eqref{eq:a}--\eqref{eq:xi}. Indeed the relationship is linear in the simulation, with the constants of proportionality ($A$) in the form 
\begin{equation}
    \label{eq:Axt}
    x^5(t) - x^5_{0} = A_x^5 t
\end{equation} measured to be 
\begin{align}
    \label{eq:A}
    \text{Run i: }
    A_{\tilde{a}}^{(\sigma)} = 0.021, 
    A_{e}^{(\mu)} = 0.014,
    A_{i}^{(\mu)} = 0.007.
\end{align}

The corresponding model is plotted with dashed lines throughout panels in Fig.~\ref{fig:aei}, which reasonably fit the simulation. We also fitted a model in which we allowed the power law index to depart from 0.20, finding a best-fit index of 0.22. Despite this noticeable deviation, the analytical scaling relations provide an overall reasonably close match to the N-body simulations.

\subsection{Evolution of radial and vertical profiles}
Given the Gaussian $\tilde{a}$ distribution and Rayleigh $e$ distribution of an initially thin ring, it is possible to predict the shape and evolution of its radial profile, which is a quantity closer to what can be directly observed. \citet[Eq.~32]{Rafikov2023} derived an expression for the radial profile given a semimajor axis distribution, $f(a)$, and a Rayleigh eccentricity distribution with scale parameter $e_\sigma$,
\begin{equation}
    \label{eq:aekernel}
    \Sigma(r) \propto \int_{r/2}^{\infty} \frac{f(a) \mathrm{e}^{-\kappa^2 / 2 e_\sigma^2}}{a e_\sigma (1 - \mathrm{e}^{-e_\sigma^2/2})} 
    \mathrm{erf} \left( \sqrt{\frac{1 - \kappa^2}{2 e_\sigma^2}} \right) \mathrm{d}a,
\end{equation}
where $\kappa^2 = \left(1 - r/a\right)^2$ and erf is the Gaussian error function. 

Plugging in a Gaussian probability density function for $f(a)$, we numerically integrated the expression, finding that the resulting radial profile is well-approximated by a Gaussian with 
\begin{equation}
    \label{eq:rae}
    \tilde{r}_\sigma = \left( \tilde{a}_\sigma^2 + e_\sigma^2 \right)^{1/2},
\end{equation}
where
\begin{equation}
    \tilde{r} = \frac{\Delta r}{r_\mu} = \frac{r - r_\mu}{r_\mu},
\end{equation}
and $\tilde{r}_\sigma$ is the Gaussian standard deviation of $\tilde{r}$ among all planetesimals. 

\begin{figure}
    \centering
    \includegraphics[width=1\linewidth]{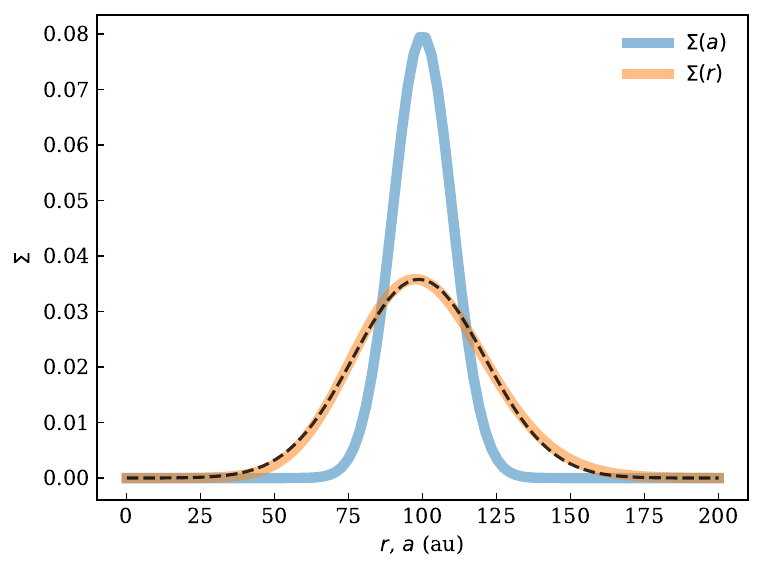}
    \caption{The theoretical radial profile resulting from a Gaussian semimajor axis distribution and Rayleigh eccentricity distribution is well-approximated by a Gaussian. Thick solid lines show the radial density profile and semimajor axis distribution, whereas the thin dashed line shows a Gaussian fitted to the radial profile. }
    \label{fig:convolve_ae}
\end{figure}

An example is shown in Fig.~\ref{fig:convolve_ae}, in which the standard deviation of the radial profile approximated with Eq.~\eqref{eq:rae} is consistent with that of the best-fit Gaussian within $1\%$. This is in fact the approximation assumed in Eq.~\eqref{eq:n}. If all constant factors were kept in Sec.~\ref{sec:scaling}, assuming a uniform number density within the radial and vertical FWHM under equipartition dispersion ratios (Eq.~\ref{eq:aei}) and $k_i = 1/\sqrt{3}$ (Eq.~\ref{eq:k}) would have predicted near exact $A$ constants (within 1\%) compared to values found by N-body simulations in Eq.~\eqref{eq:A}. Dropping all constants would still have resulted in the same $A$ values within a factor of 2. 

The vertical profile is also expected to be Gaussian, since a Rayleigh probability density function integrates to a half-Gaussian cumulative density function, considering bodies at random phases in their orbit. A full derivation is presented in \citet{Matra2019}, from which the scale height and Rayleigh scale parameter of $i$ are related by
\begin{equation}
    \label{eq:hi}
    h_\sigma = i_\sigma,
\end{equation}
where
\begin{equation}
    h = \frac{\Delta z}{r_\mu} = \frac{z}{r_\mu},
\end{equation}
and $h_\sigma$ is the Gaussian standard deviation of $h$ among all planetesimals. 

Observationally, it is more common to characterise the radial fractional width based on the FWHM \citep{Matra2025, Han2025}. Based on the relations in Eq.~\eqref{eq:aei}, \eqref{eq:rae} and \eqref{eq:hi}, $r_\mathrm{FWHM}$ and $h_\sigma$ at equipartition are conveniently related by
\begin{equation}
    \label{eq:rh}
    \tilde{r}_\mathrm{FWHM} = 10 \ h_\sigma.
\end{equation}

\begin{figure}
    \centering
    \includegraphics[width=1\linewidth]{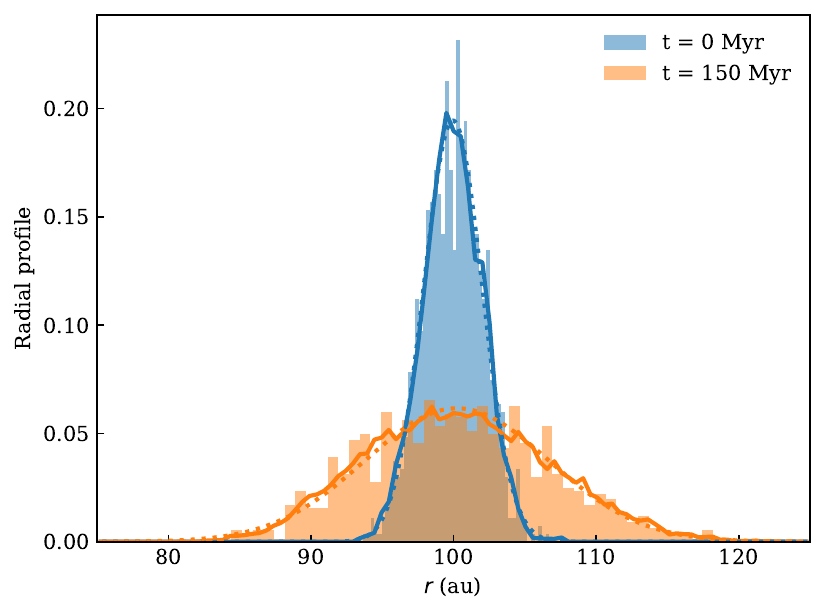}
    \includegraphics[width=1\linewidth]{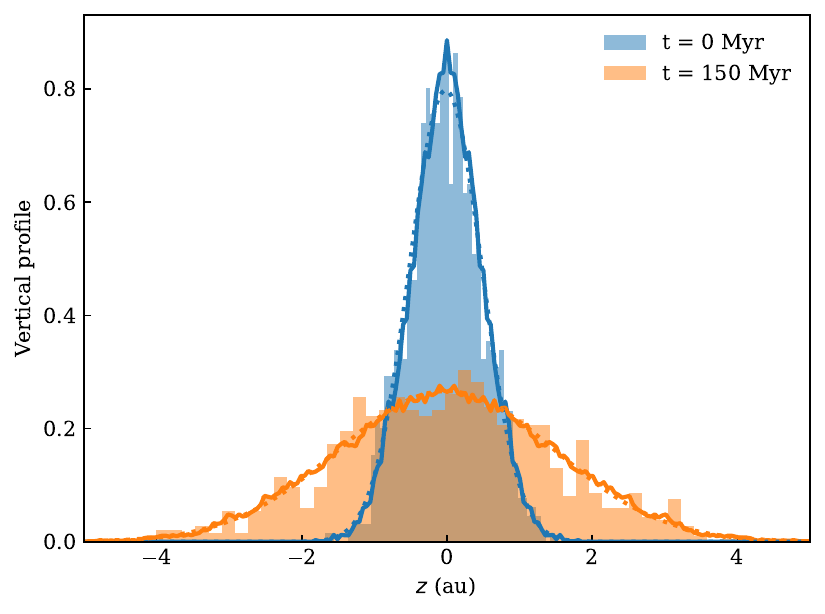}
    \caption{The radial and vertical profile at 0 and 150\,Myr in Run i. The histograms show the spatial distribution of planetesimals at a snapshot in time, whereas the solid lines show the orbit-averaged density distributions. The dotted lines are the best-fit Gaussians to the histograms. }
    \label{fig:radial_vertical_hist}
\end{figure}

\begin{figure}
    \centering
    \includegraphics[width=1\linewidth]{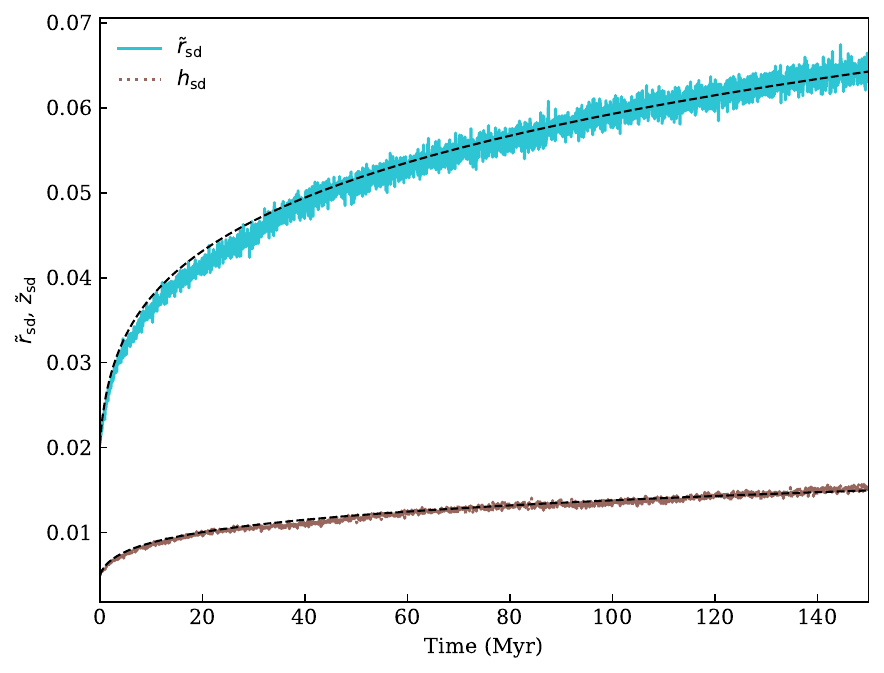}
    \caption{The evolution of $\tilde{r}$ and $h$ in Run i. Dashed lines show the $t^{1/5}$ dependence in Eq.~\eqref{eq:xi} scaled to the best-fit constants. }
    \label{fig:radial_vertical_evolution}
\end{figure}

These expectations are met by simulations. The radial and vertical profiles at the beginning and end of Run i are shown in Fig.~\ref{fig:radial_vertical_hist}, in which they are indeed well-fit by Gaussians. The evolution of the radial width and scale height are shown in Fig.~\ref{fig:radial_vertical_evolution}, which are both reasonably described by a $t^{1/5}$ dependence, with $\tilde{r}_\mathrm{FWHM} / h_\sigma = 10.1$ on average over the duration of the simulation. Analogous to Eq.~\eqref{eq:Axt}, we measure the constants of proportionality for $r$ and $h$ to be
\begin{align}
    \label{eq:Arh}
    \text{Run i: }
    A_{\tilde{r}}^{(\mathrm{FWHM})} = 0.056, 
    A_{h}^{(\sigma)} = 0.0055,
\end{align}
which indeed follow the expected equipartition relations in Eq.~\eqref{eq:rae} and \eqref{eq:rh}. 

\subsection{Scaling with disk and stellar properties}
Having investigated the $t$ dependence with Run i, we now turn to other dependencies in Eq.~\eqref{eq:xi} by comparing simulations that alter one or more parameters relative to Run i. These simulations are summarised in Table~\ref{tab:nbody}. 

\begin{figure}
    \centering
    \includegraphics[width=1\linewidth]{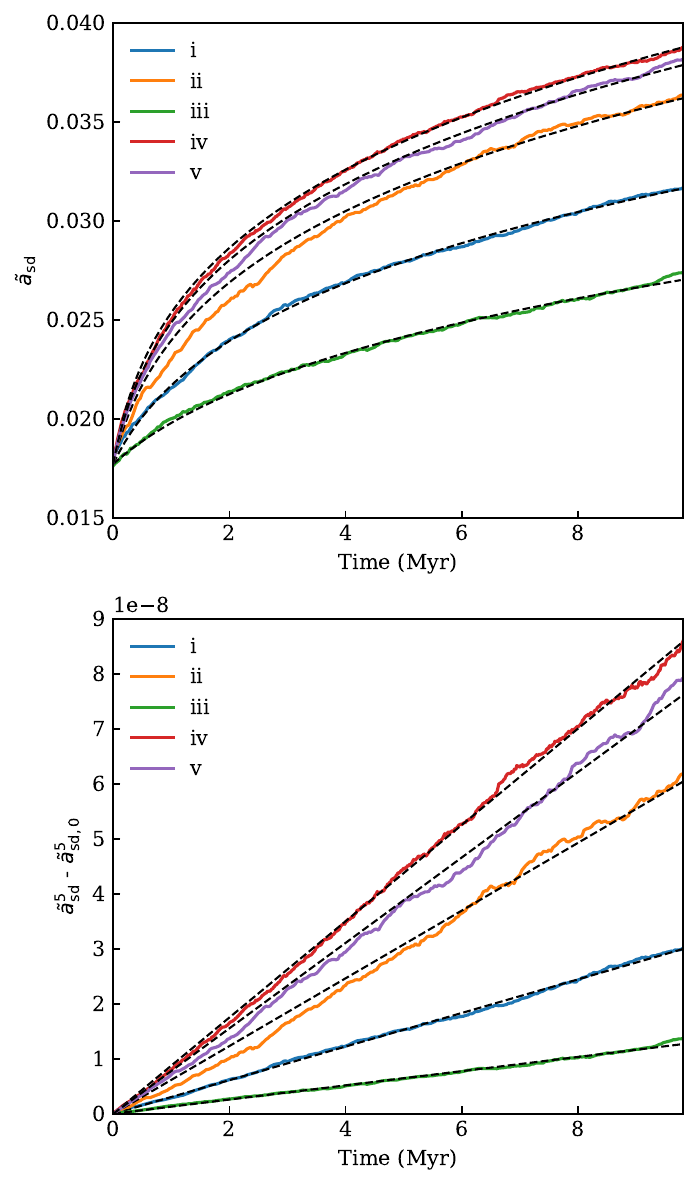}
    \caption{The evolution of $\tilde{a}$ in Runs i--v on a linear scale (top panel) and transformed according to Eq.~\eqref{eq:a} with the theoretical relation scaled to the best-fit constants (bottom panel). The best-fit constants were used to test the scaling in Eq.~\eqref{eq:scaling} by comparing Run i with other runs that alter different disk and stellar properties. }
    \label{fig:scaling}
\end{figure}

We examined the evolution of $\tilde{a}$, $e$ and $i$ between 0 and 10 Myr across all runs, with the example of $\tilde{a}$ evolution plotted in Fig.~\ref{fig:scaling}. For each simulation, we fitted the constants of proportionality between $\left( x^5 - x_0^5 \right)^{1/5}$ and $t$, where $x = \tilde{a}$, $e$ or $i$, which we used to measure the power-law scaling of $\tilde{a}$, $e$ and $i$ with respect to the number of planetesimals, the planetesimal mass, the stellar mass, and the disk's semimajor axis. 

Writing $\xi$ as
\begin{equation}
    \label{eq:scaling}
    \xi(t) = N^\nu m^\mu M_*^\gamma a^\alpha t^\tau,
\end{equation}
the analytic scaling in Eq.~\eqref{eq:xi} is equivalent to $\nu = 0.2$, $\mu = 0.4$, $\gamma = -0.3$, $\alpha = -0.3$ and $\tau = 0.2$. 
The indices measured from the simulations, averaged over $\tilde{a}$, $e$ and $i$, are $\nu = 0.21$, $\mu = 0.42$, $\gamma = -0.27$ and $\alpha = -0.33$ (fixing $\tau = 0.2$), which overall are similar to expectations based on scaling arguments.

Collecting the analytical scaling and N-body constants measured above, the orbital element evolution in equipartition is described by
\begin{align}
    \label{eq:an}
    &\tilde{a}_\sigma(t) \approx \left[ \tilde{a}_\sigma^5(t) - \tilde{a}_{\sigma, 0}^5 \right]^{1/5} = 0.021 \, \xi(t),\\
    \label{eq:en}
    &e_\mu(t) \approx\left[ e_\mu^5(t) - e_{\mu, 0}^5 \right]^{1/5} = 0.014 \, \xi(t),\\
    \label{eq:in}
    &i_\mu(t) \approx\left[ i_\mu^5(t) - i_{\mu, 0}^5 \right]^{1/5} = 0.007 \, \xi(t),
\end{align}
with underlying random velocities
\begin{align}
    \label{eq:vn}
    &\tilde{v}_\mu(t) \approx \left[ \tilde{v}_\mu^5(t) - \tilde{v}_{\mu, 0}^5 \right]^{1/5} = 0.012 \, \xi(t),\\
    \label{eq:vswn}
    &\tilde{v}_{\mathrm{swm}, \mu}(t) \approx \left[ \tilde{v}_{\mathrm{swm}, \mu}^5(t) - \tilde{v}_{\mathrm{swm}, \mu, 0}^5 \right]^{1/5} = 0.018 \, \xi(t),
\end{align}
and corresponding observational quantities evolving according to
\begin{align}
    \label{eq:rn}
    &\tilde{r}_\mathrm{FWHM}(t) \approx \left[ \tilde{r}_\mathrm{FWHM}^5(t) - \tilde{r}_{\mathrm{FWHM}, 0}^5 \right]^{1/5} = 0.056 \, \xi(t),\\
    \label{eq:hn}
    &h_\sigma(t) \approx\left[ h_\sigma^5(t) - h_{\sigma, 0}^5 \right]^{1/5} = 0.0055 \, \xi(t),
\end{align}
where
\begin{align}
    \label{eq:xin}
    \begin{split}
    \xi(t) = {} &
        \left( \frac{N}{1024} \right)^{1/5}
        \left( \frac{R}{1000 \ \mathrm{km}} \right)^{6/5}
        \left( \frac{\rho}{2.7 \ \mathrm{g}\,\mathrm{cm}^{-3}} \right)^{2/5}\\
        & \times \left( \frac{M_*}{1 \ M_\odot} \right)^{-3/10}
        \left( \frac{a_\mu}{100 \ \mathrm{au}} \right)^{-3/10}
        \left( \frac{t}{1 \ \mathrm{Myr}} \right)^{1/5},
    \end{split}
\end{align}
and the approximation is valid when the orbital element or observed quantity $x$ satisfies $(x / x_0)^5 \gg 1$. 

\section{Application to disk observations} \label{sec:observations}
The aim of this section is to investigate the extent to which the observed differences in ring widths between protoplanetary and debris disks could be plausibly attributed to intrinsic ring broadening via gravitational kicks between its constituent planetesimals.

\subsection{Protoplanetary and debris disk sample}
Here we focus on millimetre-wavelength observations of protoplanetary and debris disks primarily with ALMA, which are sensitive to dust grains sufficiently large that we can ignore dynamical effects of stellar radiation on the observed disk structure \citep{Milli2026, Jankovic2026, Han2026b}. 
For protoplanetary disks, we used the compilation of protoplanetary ring radius and their FWHM by \citet{Bae2023}, which incorporates data from multiple observing programs such as DSHARP \citep{Andrews2018}. For debris disks, we used the REASONS sample \citep{Matra2025} of millimetre-resolved debris disks. A subset of disks within REASONS were observed by ALMA at higher resolution as part of the ARKS program \citep{Marino2026}, which we used to replace the corresponding REASONS data for this analysis. 

Note that many protoplanetary and debris disks contain multiple rings and we treat each ring as an independent data point. The \citet{Bae2023} sample provides measurements of individual rings in protoplanetary disks. While the REASONS sample does not fit to individual rings within multi-ring debris disks, all 5 such disks are part of the ARKS compilation which does fit to each ring independently, and we use the nonparametrically measured values of the statistially significant rings in the ARKS sample \citep{Han2026}. We also excluded REASONS disks for which only an upper limit can be placed on the radial FWHM of the ring. 

The corresponding studies on these datasets provide details on the sample selection and potential biases. In general, the most obvious bias of the resulting sample is towards big and bright disks. For debris disks especially, such systems are more common around younger stars and earlier type stars (e.g., A-type), however despite this bias the sample covers a wide range of stellar ages and luminosities.

\subsection{Radial width evolution}

\begin{figure}
    \centering
    \includegraphics[width=1.0\linewidth]{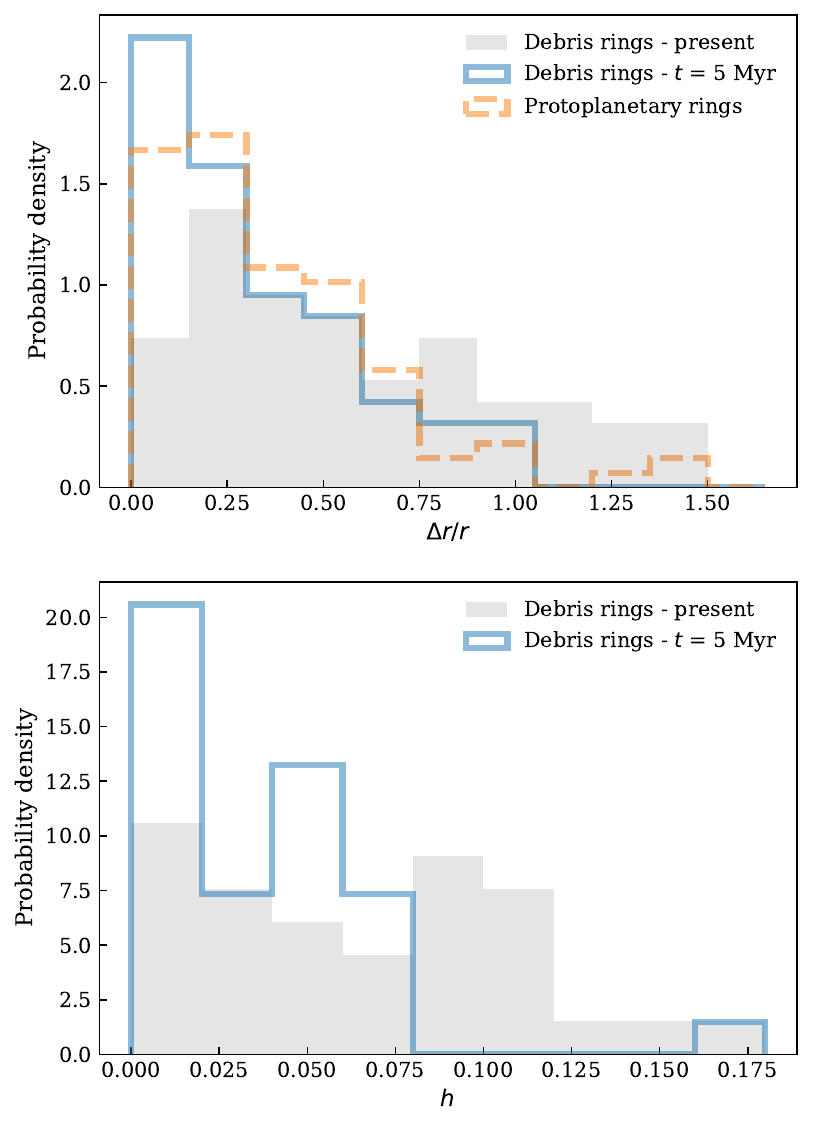}
    \caption{The fractional radial FWHM ($\Delta r / r$) and scale height ($h = h_\sigma$) distribution of planetesimal belts as measured from individual rings in debris disk \citep{Matra2025, Marino2026, Han2026} and the radial fractional FWHM distribution of rings in protoplanetary disks \citep{Bae2023} from ALMA observations. Given the age of each debris disk, we evolved the width of the rings back in time based on Eq.~\eqref{eq:rn} and plotted their radial and vertical width distributions had they been observed at $t = 5$\,Myr after initial formation in a narrow ring. The resulting time-evolved fractional width distribution is similar to that of protoplanetary rings from which they may have formed. }
    \label{fig:histogram}
\end{figure}

The REASONS sample suggested that the median fractional width ($\tilde{r}_\textrm{FWHM} = \Delta r / r$) of debris disks is 0.71 \citep{Matra2025}, which is over twice as large as rings in resolved protoplanetary disks at 0.29 \citep{Bae2023}. 
It was not clear whether this difference is real or due to the often lower resolution used to observe the fainter debris disks compared to protoplanetary disks, thereby being unable to resolve individual rings in multi-ring disks, or the accurate width of marginally resolved single-ring disks. 

The recent ARKS ALMA program \citep{Marino2026} reached significantly higher resolution, enabling a more robust characterisation of debris disk ring widths. This newer dataset suggests that the median fractional width among the high-resolution debris disk sample is in fact only 0.33 \citep{Han2026}, however a sizable fraction of the debris ring population is still measured to be wider that protoplanetary rings. 

Several possibilities have been suggested to explain why the structure of debris rings may not be inherited directly from those of protoplanetary rings. These include scenarios in which some planetesimals belts formed wide in the protoplanetary disks, where protoplanets drive dust trap migration, leaving planetesimals formed over a wider range of radii than the ALMA rings observed at any instant in time \citep{Miller2021}; ones in which a migrating photoevaporation front during protoplanetary disk dispersal triggers planetesimal formation over a wide range of radii \citep{TLau2025, LiChiang2026}; and ones in which planetesimal belts that have formed are scattered over time by migrating planets, similar to the influence of Neptune on the Solar System's Kuiper belt \citep{Gomes2005, Matra2025}. 

To understand the demographic data from these recent observations, here we are motivated by a simple disk evolution scenario without invoking external perturbers such as planets. Assuming that protoplanetary rings observed by ALMA represent the initial site of planetesimal formation, we seek to test whether the $\Delta r/r$ difference between protoplanetary and debris rings could be consistent with the instrinsic self-scattering among planetesimals by virtue of them having mass. 

Assuming that planetesimal belts initially formed narrow, and that the equilibrium belt width has not been reached at the time of observation of debris disks (an assumption which we verify in Sec.~\ref{sec:mass}), we used Eq.~\eqref{eq:rn} to evolve observed debris rings back in time, obtaining the fractional width distribution of planetesimal belts around the class II protoplanetary disk phase. The disk ages assumed are those summarised in \citet{Marino2026} for ARKS disks and \citet{Matra2025} for REASONS disks, with references for the age provided therein. The choice of time of initial planetesimal formation is somewhat uncertain, however planetesimal formation is likely well underway $\lesssim 0.5$\,Myr during or before the class II protoplanetary disk phase \citep{SeguraCox2020, Hsieh2025}, and that self-stirring may have been ongoing even in the presence of gas. Considering the approximate planet formation and protoplanetary disk dispersal timescales of a few Myr \citep{Williams2011, Pfalzner2022}, and isotopic evidence from the Solar System that suggest planetesimal formation started within the first Myr (CAIs) and extended into the next $\sim$3\,Myr (CCs, \citealp{Kruijer2017}), we evolved the observed debris rings to $t = 5$\,Myr for comparison with the protoplanetary ring population. 

\begin{figure}
    \centering
    \includegraphics[width=1.0\linewidth]{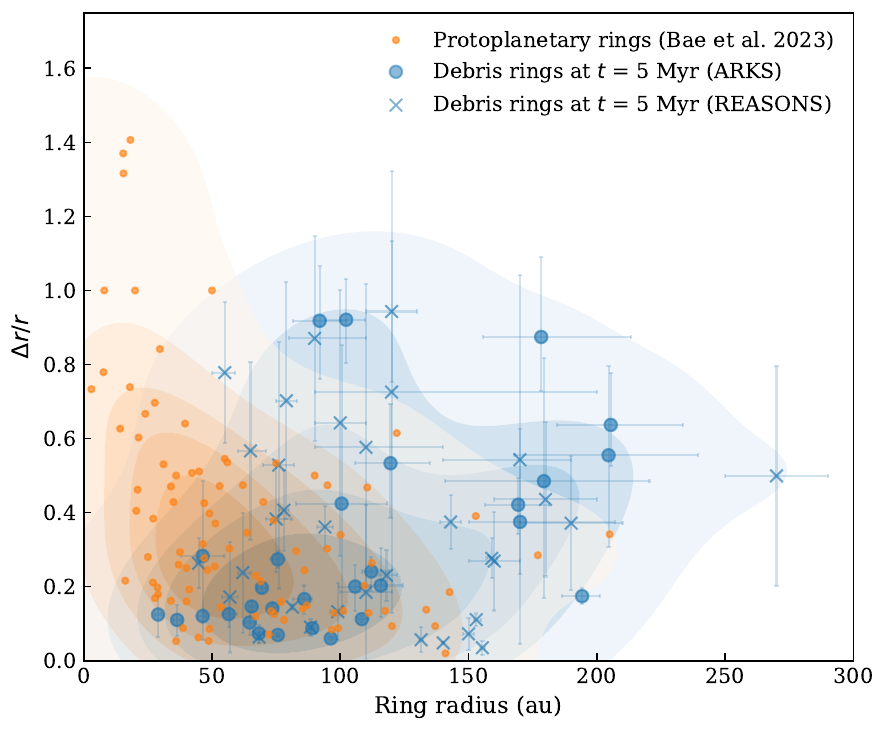}
    \caption{The time-evolved fractional widths of debris rings at 5\,Myr assuming formation in a narrow ring (see Fig.~\ref{fig:histogram}) as a function of ring radius based on ALMA observations from the ARKS \citep{Marino2026} and REASONS \citep{Matra2025} datasets. Rings in protoplanetary disks \citep{Bae2023} are plotted for comparison. This figure can be compared to Fig.~5 in \citet{Matra2025}, which shows the present distribution of REASONS debris rings without playing them back in time.}
    \label{fig:rdr}
\end{figure}

Fig.~\ref{fig:histogram} shows the distribution of protoplanetary rings and of debris rings at present and inferred at 5\,Myr. The debris rings played back in time show strong resemblance to the protoplanetary ring distribution, skewing heavily towards the narrow end. Fig.~\ref{fig:rdr} provides further details on the rings radius and fractional width distribution, which can be compared with Fig.~5 in \citet{Matra2025} for the REASONS dataset based on measured values at present. In Fig.~\ref{fig:rdr}, the kernel density estimations overlap along the $\Delta r / r$ dimension. The larger radius in debris rings could reflect the observational bias towards more luminous stars, which is correlated with more radially extended disks \citep{Matra2018, Matra2025}. 

Taken together and with the caveats described above (and summarised in Sec.~\ref{sec:limitations}), the similarity between time-evolved debris rings and protoplanetary rings suggests that radial broadening due to viscous stirring, taking into account both the semimajor axis and eccentricity evolution, can offer a plausible general explanation for the debris ring population to appear wider than protoplanetary rings. 

\subsection{Mass of planetesimals in debris disks}
\label{sec:mass}

With knowledge of the stellar mass, ring radius and age, the viscous stirring rate in Eq.~\eqref{eq:xin} predicts the product of the total ring mass ($M_\mathrm{disk}$) and the mass of each individual planetesimal within the ring ($m$), $M_\mathrm{disk} \times m$, assuming that the rings formed narrow and evolve under self-stirring. Although this $Nm^2$ dependence implies that the total disk mass cannot be uniquely constrained, this product is still an illuminating quantity for understanding the kind of planetesimals rings that protoplanetary disks are able to form. 

To measure $mM_\mathrm{disk}$, we assumed belt radii, stellar ages and stellar masses provided in the corresponding datasets (\citealp{Matra2025, Marino2026}, and references therein). As the REASONS dataset did not provide stellar masses, we estimated the stellar mass as $(M_*/M_\odot) = (L_*/L_\odot)^{0.25}$ using the $L_*$ values provided in the dataset (except for Vega which, due to its proximity, is well-determined to deviate from this scaling as a result of its spin and projection angle, so we used the mass estimate from \citealt{Monnier2012}). The ARKS stellar mass modelling returned values typically within 10\% from this scaling across the sample, so we assumed an uncertainty of 10\% for the REASONS stellar masses obtained from this scaling. 

\begin{figure}
    \centering
    \includegraphics[width=1.0\linewidth]{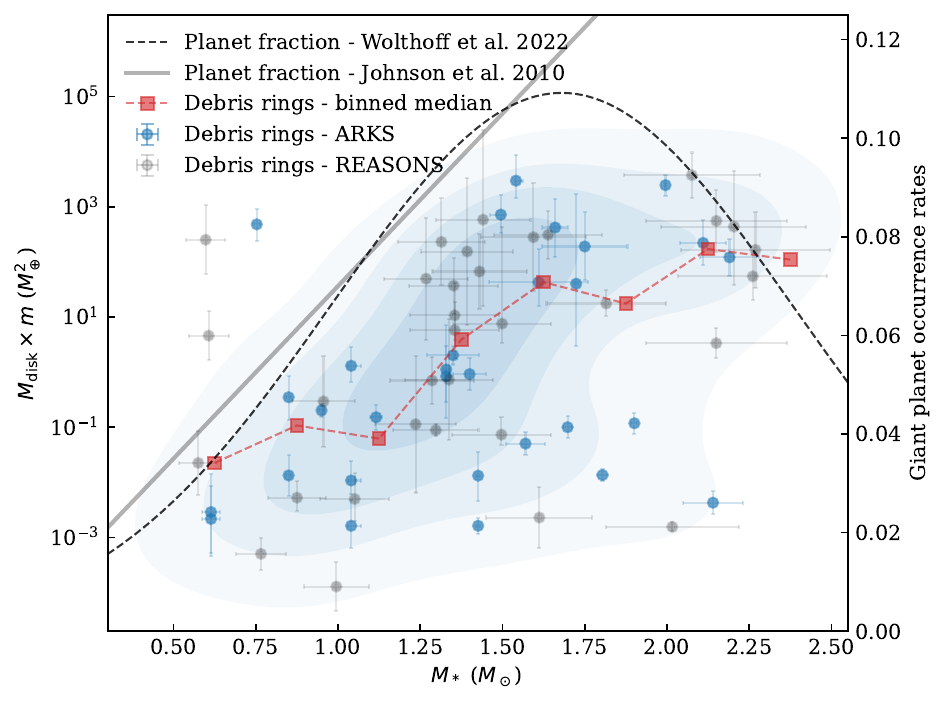}
    \caption{The product of the total debris ring mass ($M_\mathrm{disk}$) and the mass of each constituent planetesimal ($m$). This $M_\mathrm{disk} \times m$ estimate comes from the rate of radial broadening required for the planetesimal belt to reach the observed fractional width at the age of the system. The distribution appears to be skewed towards higher masses for more massive stars, which can be described by an envelope similar to the occurrence rate of giant planets as a function of stellar mass. The planet fractions plotted come from \citet{Johnson2010} and \citet{Wolthoff2022}, which attempted to correct for observational biases. The median $M_\mathrm{disk} \times m$ in each $0.25\,M_\mathrm{\odot}$-wide stellar mass bin is overplotted.}
    \label{fig:mass}
\end{figure}

The $mM_\mathrm{disk}$ measurements based on these assumptions are shown in Fig.~\ref{fig:mass}. These mass products span a large range and extend across 6 orders of magnitude. The smallest $mM_\mathrm{disk}$ values are $\sim10^{-3} \, M_\oplus^2$. In order for $M_\mathrm{disk}$ not to exceed the maximum debris disk mass of $\sim$100\,$M_\oplus$ (assuming a protoplanetary disk gravitational stability limit of $0.1 \, M_*$ and a dust to gas ratio of 1\%), the minimum radius of the constituent ``large bodies'' is $\sim100$\,km. Decreasing the $M_\mathrm{disk}$ assumption by a factor $\chi$ would imply a $\chi$ times larger minimum mass for largest planetesimals.

The largest $mM_\mathrm{disk}$ values across these planetesimal rings are $\sim10^3 \, M_\oplus^2$. For a disk mass of 100\,$M_\oplus$, this would require stirring bodies on the order of $10 \, M_\oplus$, corresponding to a disk of small bodies being stirred by $\sim$10 super-Earths. 

The majority of disks appear to lie between these two extremes. The median $mM_\mathrm{disk}$ is 1.3\,$M_\oplus^2$. Such a product could be achieved by, for instance, a disk of 100 Mars-sized bodies.

There is in general considerable uncertainty surrounding the size of the largest planetesimals in debris disks (\citealp{Krivov2021, Zawadzki2026, Chiang2026}; Jankovic et al. submitted). Our own Kuiper belt's largest bodies are on the order of 1000\,km in radius (e.g., Pluto). Although these objects are few in number, it is thought that the Kuiper belt used to be two to three orders of magnitude more massive before being scattered by Neptune under the Nice model \citep{Levison2008}. To explain the under-abundance of resonant planetesimals found in the Kuiper Belt had Neptune migrated smoothly, a few thousand Plutos in the early Solar System have been invoked to induce grainy planetesimal-driven migration trajectories for Neptune, corresponding to a disk mass on the order of $\sim$10\,$M_\oplus$, albiet with large uncertainties \citep{Nesvorny2018}. Such a scenario would correspond to a $mM_\mathrm{disk}$ of $\sim0.03 \, M_\oplus^2$, which is indeed close to the median for a 1\,$M_\odot$ star in Fig.~\ref{fig:mass}. 

Based on these $mM_\mathrm{disk}$ values inferred, we verified that the scaling in Eq.~\eqref{eq:xin} used is applicable by checking that the velocity dispersion required in each disk is lower than the escape velocity, since the deflection radius shrinks to the physical radius of the planetesimals when the velocity dispersion reaches the escape velocity, thus ceasing to grow. 
Assuming $M_\mathrm{disk}$ equal to the maximum disk mass of 100\,$M_\oplus$, which would necessitate the lowest possible $m$ and thus $v_\mathrm{esc}$, we find that this condition still holds for all rings across the sample, with half the sample required to have a $v_\mathrm{swm}$ below $0.2 \, v_\mathrm{esc}$.

Following from Eq.~\eqref{eq:vswn}, the timescale required for $v_{\mathrm{swm}, \mu}$ to be stirred up to $v_{\mathrm{esc}}$ is
\begin{equation}
    \label{eq:eq}
    \begin{split}
        t_\mathrm{eq} = {} & 600 \, \mathrm{Myr}
        \left( \frac{N}{1024} \right)^{-1}
        \left( \frac{R}{1000 \ \mathrm{km}} \right)^{-1} \\
        & \times \left( \frac{\rho}{2.7 \ \mathrm{g}\,\mathrm{cm}^{-3}} \right)^{0.5}
        \left( \frac{M_*}{1 \ M_\odot} \right)^{-1}
        \left( \frac{a_\mu}{10 \ \mathrm{au}} \right)^{4}.
    \end{split}
\end{equation}
We find for 75\% of the sample, the age of these disks is at most 1\% of $t_\mathrm{eq}$ under these $M_\mathrm{disk}$ assumptions, with the velocity dispersion still expected to be growing across all disks. 

Under the viscous stirring interpretation, a few conclusions can thus be drawn. Firstly, the planet formation outcome as told by debris rings is highly diverse. When measured by $\sqrt{m M_\mathrm{disk}}$, the mass of bodies that form span 3 orders of magnitude. Even within a given stellar mass bin, a wide range of planetesimal belt masses appear to emerge from the protoplanetary disk at 10s of au and beyond. 

Secondly, the distribution of $m M_\mathrm{disk}$ appears to show dependence on the stellar mass, with $m M_\mathrm{disk} > 1\,M_\oplus^2$ disks being markedly more abundant for stars more massive than $\sim$1.2\,$M_\odot$, i.e., A and early F stars. Interestingly, the envelope of this trend is rather similar to the giant planet occurrence rate, which also peaks at around 1.5 to 2\,$M_\odot$. If it is true that planetesimals form in initially narrow rings before spreading out via mutual scattering, these inferred debris disk masses could indeed be the surviving record that the final ingredients of giant planetary cores are more abundant around A and F stars even at $\sim$100\,au, in synchrony with the frequency of giant planets that do eventually form and are detected typically interior to these disk radii. The large scatter within this envelope could hint at wide variations in either the efficiency of solid body coagulation, or in whether the solid mass in protoplanetary disks end up in planets or the debris disk.

Relatedly, it is an interesting question as to what the largest $m M_\mathrm{disk}$ values represent. A handful of super-Earths having grown oligarchically at these locations could be one plausible scenario \citep{Kokubo1998, Zang2025}. How these super-Earths could continue to evolve is worth considering, as they could either rapidly stir up the ring of any remaining smaller planetesimals and themselves risk mutual chaotic ejection, or if they end up being sufficiently spaced out, they could remain within the disk and continue to stir it as we observe the disk today. The Hill radius of a 10\,$M_\oplus$-body at 100\,au is approximately 2\,au. If these bodies were to be separated by 5 times the Hill radius \citep{Kokubo1998}, a handful of such bodies in the disk would then occupy a radial range of tens of au, which is comparable to the widest debris disks observed with $\Delta r / r \gtrsim 1$. Any gaps in the disk carved by the chaotic overlap of mean motion resonances would have a semimajor axis width of $\sim$15\,au \citep{Morrison2015}, which would typically be unresolved with the majority of current ALMA observations of such wide disks \citep{Han2026}, even if these gaps were not partially filled by highly eccentric planetesimals and dust. 

Alternatively, it is possible that these super-Earth stirrers once existed in the disk and scattered its smaller planetesimals, but have since either migrated away, or the disk might be stirred by an external planet that was never embedded within the ring. In any of these scenarios, perturbation by massive bodies distinct from Pluto-sized planetesimals is likely to have occurred, suggesting that large stirrers influencing the tens of au region and beyond are more commonly found around higher-mass stars.

Finally, the fact that even the lowest plausible $m$ estimates correspond to an equilibrium radial width that is wider than observed suggests that virtually no debris disk has reached equilibrium through viscous stirring alone, and that these observed disks are likely continuing to broaden over time both radially and vertically. We return to this point in Sec.~\ref{sec:disk_discussion}.

\subsection{Scale height evolution}
\label{sec:h_evolution}

\begin{figure}
    \centering
    \includegraphics[width=1.0\linewidth]{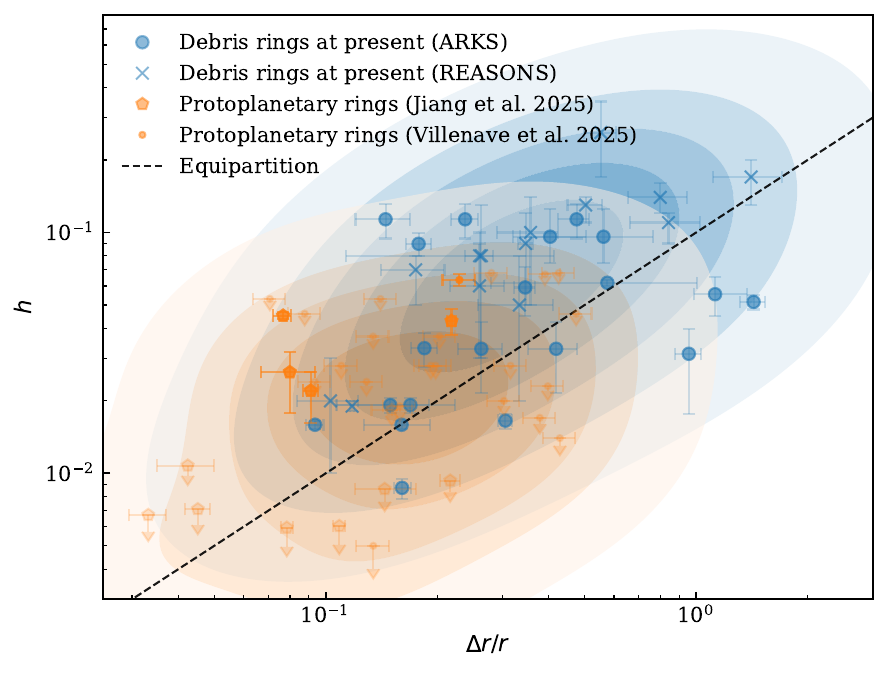}
    \caption{The scale height and fractional radial width of protoplanetary and debris rings. 
    The dashed line shows the $h_\sigma = 0.10 \, \Delta r_\mathrm{FWHM} / r$ relation (Eq.~\ref{eq:rh}) expected at equipartition between the semimajor axis, eccentricity and inclination dispersions of an initially narrow ring, which the population of resolved rings in both protoplanetary and debris disks lie close to. 
    Kernel density estimations are plotted to provide visual guidance of their distributions. Given the prevalence of upper limits in the protoplanetary disk sample, we arbitrarily assumed the upper limits to be 50\% larger than the true value when drawing these contours.
    }
    \label{fig:hr}
\end{figure}

While we have focused on the radial broadening of planetesimal rings so far, analogous vertical thickening is expected to occur in synchrony (Fig.~\ref{fig:snapshots}). Using ARKS (values fitted with the \texttt{frank} code, \citealt{Jennings2020, Terrill2023, Zawadzki2026}) and REASONS measuerments of debris disk scale heights with both upper and lower constraints, we applied the time evolution scaling of Eq.~\eqref{eq:hn} and plot their scale height distribution both as presently observed and evolved back to $t = 5$\,Myr in Fig.~\ref{fig:histogram}. Whereas the median $h_\sigma$ for debris rings is measured to be 0.061 across the combined dataset, the inferred $t = 5$\,Myr median is only 0.028. 

A scale height comparison between protoplanetary and debris disks is more challenging than for the radial structure, which is mainly due to the lack of well-constrained dust scale heights for a substantial protoplanetary disk sample, with studies more commonly yielding only upper limits on $h$ \citep{Villenave2025, Jiang2025}. Nonetheless, we attempted to draw some comparison based on these upper limits. We collected protoplanetary disk dust scale height constraints determined by \citet{Villenave2025} of targets that have corresponding ring radius and ring width measurements summarised in \citet{Bae2023}. We assumed uncertainties on $\Delta r / r$ to be 10\% for protoplanetary rings. For 6 of the disks from the DSHARP ALMA program \citep{Andrews2018}, modelling by \citet{Jiang2025} simultaneously constrained the ring radius, ring width and scale height, so we use those values instead. 

We plotted the resulting protoplanetary ring sample in Fig.~\ref{fig:hr} alongside debris rings. Interestingly, the distribution of $h$ and $\Delta r / r$ for both protoplanetary and debris rings on a population level lie close to the equiparition relation between the radial width and scale height in Eq.~\eqref{eq:rh}. Most debris rings are consistent with equipartition within a factor of 2 (accounting for 1$\sigma$ uncertainties), with the vast majority ($\sim$90\%) lying within a factor of 3 from equipartition, suggesting that planetesimal belts around main-sequence stars may in general be in or close to equipartition. Overall, debris rings as a population show a slight enhancement in $h$ relative to $\Delta r / r$ in comparison to semimajor axis--eccentricity--inclination equipartition. Limitations in the resolution of observations could contribute to such an offset, as there exist examples in which resolution improvements have led to smaller scale height measurements \citep{Zawadzki2026}. 

Given the natural tendency for $a$, $e$ and $i$ to viscously spread as studied in Sec.~\ref{sec:scaling}, even a ring of planetesimals that formed over a (almost infinitesimally) narrow range of semimajor axes will exhibit a $\Delta r/r$ that reaches 10 times $h$. A ring that formed over a wider range in $a$ would naturally exhibit an even wider $\Delta r/r$ relative to $h$ than achievable with viscous stirring alone, unless $i$ is formed in a way that conincidentally matches the $a$ dispersion for each system. The fact that the observed population of planetesimal rings across a wide range of planetary system ages exhibits $\Delta r/r$ close to (or even smaller than) 10 times $h$ would thus appear to argue against planetesimal formation over a wide range of semimajor axes as the main formation channel, instead favouring planetesimal formation in narrow rings. 

As noted above, observational limitations could mean that the true $h$ values are smaller than those cited here. However, the debris disks observations considered here are taken from a vetted sample which removes measurements with large uncertainties or those with only upper limits on $h$, and are modelled by procedures designed to account for noise and beam size effects \citep{Matra2025, Marino2026, Han2026, Zawadzki2026}. Naively, it thus appears unlikely that the true $h$ across the population is an order of magnitude or more smaller than what has been inferred from observations. Examples of planetesimal belts distantly removed from equitation, either with $\Delta r / r$ on par with $h$, or with $\Delta r / r$ 100 times larger than $h$, are thus expected to be limited in number. Sec.~\ref{sec:limitations} discusses further limitations relating to these inferences.

\section{Discussion} \label{sec:discussion}

\subsection{Evolution of ringed substructures from protoplanetary to debris disks} \label{sec:disk_discussion}

We considered a picture in which planetesimals form in initially narrow rings, similar to those observed in $\sim$Myr-old class II protoplanetary disks but possibly at even earlier stages, before they scatter each other into radially wider and vertical puffier rings over time, like those observed in $\sim$10\,Myr- to Gyr-old debris disks. 

From the protoplanetary disk perspective, there is observational and theoretical support for planetesimal formation in protoplanetary rings. Detailed characterisation of the dust distribution and gas velocity and pressure structure in protoplanetary disks in the exoALMA program \citep{Teague2025} have suggested that the majority of ringed substructures with such measurements supports dust rings as sites of pressure traps, where inwardly migrating millimetre-sized-dust accumulates \citep{Stadler2025}. It is theoretically expected that sufficiently high dust-to-gas ratios ($\sim$1) in these dust traps trigger two-fluid instabilities, such that millimetre-sized pebbles, which formed from sticking collisions between micron-sized dust grains inherited from the interstellar medium, rapidly collapse under gravity to form km-sized planetesimals, which go on to grow into bodies as large as thousands of km in size by accreting pebbles under the assistance of gas drag or by pairwise collisions, depending on the location and conditions of the disk \citep{Youdin2005, Johansen2007, Armitage2010}. These planetesimals are thought to be the final ingredients in assembling terrestrial planets in the inner system or the cores of giant planets in the outer system which may become large enough to accrete gas. 

From the debris disk perspective, the picture considered here is somewhat different from what is often assumed. Given that debris disks, like planets, form from protoplanetary disks, their structures are likely the result of both inhertiance from protoplanetary disks and ongoing interactions with planets. Dynamical studies that use the radial profile of debris disk to infer planetary perturbers (e.g., \citealp{Marino2018, Marino2019, Pearce2022, Han2026}) and the collisional evolution of planetesimals (e.g., \citealp{Kennedy2010, Marino2021, ImazBlanco2023}) have sometimes considered a picture involving a radially broad initial planetesimal distribution, on top of which planets are added to study how they interact with the disk. This is motivated in part by the minimum mass solar nebula and the power-law-like baseline radial profile of protoplanetary disks. 

In contrast, the picture that we study here assumes planetesimal belts to start narrow, motivated by the the ringed structures that sit atop the power-law dust baseline as sites of efficient planetesimal formation \citep{Stammler2019, Carrera2021, Jiang2023, Stadler2025, Zhao2025}. We also ignored any possible planetary perturbers, implying that any substructures are interpreted as the protoplanetary disk structure smoothed by ongoing radial and vertical spreading. Some observational evidence of this picture appears to be provided by the finding that most debris rings lie close to the equipartition relation between the radial width and vertical height for an initially narrow ring, which suggests that the initial semimajor axis range of planetesimals at formation may indeed be narrowly confined. 

Such a picture, if true with any degree of generality, would bear the perhaps pessimistic implication that the sharpness of debris disk edges reflect not the collisional erosion of an initial power law radial profile, or sculpting by planets of a certain mass and orbit alone, but instead the steady shallowing of the edges of a Gaussian radial profile that started as a narrow ring; and the ring-shaped gaps observed in debris disks not the result of sculpting by planets in an initially broad radial distribution of planetesimals via the chaotic overlap of mean motion resonances alone \citep{Morrison2015}, but what remains of gaps already present in protoplanetary disks after being smoothed by the steady advancement of planetesimal rings on either side. 

However, planets are unlikely to be fully removed from this picture, for even inheritance of gaps requires an explanation of their origin at an earlier, gas-rich stage, possibly again involving the influence of planets. Furthermore, if the perturbative reach of planets within the gas disk \citep{Zhang2018} is further than that during the debris disk phase \citep{Morrison2015, Marino2019, Han2026}, then as initially narrow rings viscously spread, a fraction of them is likely to encounter the planet's sphere of influence on the way, their edge thus becoming shaped by planet sculpting. Further complexity would be added if the planets themselves were to form within and migrate away from the planetesimal rings \citep{Jiang2023}, the inner edge of which they subsequently carve, again creating radial profiles truncated by the planet's gravitational chaotic zone \citep{Morrison2015}. 

Observationally, the diverse range of radial profile shapes observed among debris disks, many of which deviating from Gaussian by varying degrees, may be evidence that more complex processes, possibly involving planets, are required to understand the full suite of planetesimal belt substructures and their connection to the planet formation timeline and resulting planetary system architecture. The scatter among the radial width and vertical height relation may be further evidence of these additional dynamical processes at play.

\subsubsection{Ongoing width evolution in debris disks}
Our findings add some degree of complexity to interpreting disk structures when considering the suggestion that viscous spreading is likely still ongoing in virtually all debris disks, and likely for as long as the planetary system exists given the slow widening rates (Eq.~\ref{eq:eq}) at $\sim$100\,au. Even at an age of 1\,Gyr, which is towards the higher end among resolved debris disks, the scale height and radial width reached is still a factor of a few lower than the equilibrium distribution. Given the steep $a_\mu^4$ dependence, however, the equilibrium ring distribution may be reached much faster for rings at $\sim$10\,au, possibly within the age of the system. 

While we have not explicitly incorporated the distribution of these smaller steady-state dust-producing planetesimals into our model, they are expected to radially and vertically diffuse alongside the large planetesimals. In the two-population set of large and small planetesimals referenced in Sec.~\ref{sec:scaling}, the dispersion in orbital elements (and thus radial width and vertical height) of small planetesimals grow as $t^{0.25}$ under stirring by the large planetesimals \citep{Ida1992}. Since we find that the dispersion of the large planetesimals themselves grow as $t^{0.20}$, the small planetesimals will always catch up to the large planetesimals as they spread, thereby still tracing where the large bodies reside. 

The vertical height of debris disks has been used as a measure of $v_\mathrm{esc} / v_\mathrm{K}$ of the largest bodies in the disk \citep{Daley2019, Vizgan2022, Zawadzki2026}. It seems from this analysis that the distribution of the largest bodies may themselves still be evolving, thus the observed scale height is a reflection of the combined effect of $v(t) / v_\mathrm{K}$ of the largest bodies, where $v(t) < v_\mathrm{esc}$, and the viscous stirring of smaller dust-producing planetesimals under the large body distribution at that point in time. For a more reliable estimate, it may be thus necessary to take into consideration the age of the system when inferring the largest planetesimals or disk mass.

\subsubsection{(Lack of) collisions}  \label{sec:collisions}
An omission of the picture we have considered thus far is the impact of collisions. While we have treated the observed ALMA debris rings as a proxy for the distribution of planetesimals, what we are observing in reality is emission from dust and pebbles sustained by collisions between planetesimals, as the planetesimals are of much lower combined emitting area than the dust that they produce. As radiation pressure removes the smallest dust in the collisional cascade, these collisions eventually deplete the mass budget of planetesimals, causing the radial profile to become suppressed starting from smaller radii in the disk \citep{Kennedy2010, ImazBlanco2023}, where the collisional timescale is shorter. For a broad disk, this has been predicted to cause a $t^{0.2}$ growth in peak radius \citep{Wyatt2007}, which would suggest that the observed radial structure may not reflect that of planetesimals had they been evolving under gravity but without collisions. 

However, this effect may be mitigated by the fact that (1) a large fraction of rings that have been observed are relatively narrow, leaving little room for the outward propagation of the peak radius, and (2) bodies participating in the steady-state collisional cascade have been estimated to be $\sim$km in size \citep{Krivov2021}, rather than the 1000\,km-sized bodies as we have studied. Although these larger bodies may not be abundant enough to collide in steady state, they could still collide sporadically and inject fresh boulders into the steady-state cascade, much like the giant impact thought to give rise to the Pluto--Charon binary in the Solar System \citep{Canup2005}. Such impacts can act as an additional (if not primary) reservoir of debris mass \citep{Chiang2026} that are eventually captured by observations, on average restoring the outwardly propagating emission front to the location of the underlying distribution of large planetesimals. 

It is not immediately clear how sporadic collisions between these slowly viscously spreading large planetesimals would be reflected in the infrared flux of debris disk populations as a function of age. The collision rate between planetesimals is $1/t_\mathrm{col} \propto v^{-3} \propto t^{-3/5}$ from Eq.~\eqref{eq:vdvdt}, replacing the $(v_\mathrm{esc} / v)^4$ enhancement for scattering to $(v_\mathrm{esc} / v)^2$ for collisions from gravitational focussing. 
Planetesimals larger than 1\,km are bound primarily by gravity rather than material strength \citep{Stein2026}. The fragmentation velocity is then of order the escape velocity, so at equilibrium, collisions between these largest planetesimals are in general destructive. However, even while the velocity dispersion is growing, a fraction of high-velocity bodies at the Rayleigh tail will reach sufficient relative velocities to disrupt each other, and collisions in general release some fraction of the mass as debris. The dependence of the average mass released per collision on $t$ is more difficult to quantify with analytic scaling, but should increase with $t$ as relative velocities increase. The resulting time dependence of dust mass should therefore be a shallower dropoff than $1/t_\mathrm{col} \propto t^{-3/5}$. 

Models for the flux evolution of debris disks have predicted a range of time scaling depending on model assumptions, ranging between $M_\mathrm{disk}(t) \propto t^{-1}$ \citep{Wyatt2007} to $t^{-0.3}$ \citep{Lohne2008} expected from steady-state collisions between smaller planetesimals, whereas some observations have suggested an even steeper dependence in certain populations (e.g., $t^{-2}$ for the $\beta$~Pic moving group \citealp{Pawellek2021}). If the theoretical flux dropoff for smaller planetesimals in steady-state collisions is truly steeper than that from non-steady-state collisions, these observations in the $\beta$~Pic moving group could mean that while the largest planetesimals set the agenda for how the spatial distribution of the disk evolves, their collisions, even if a source of debris mass, are too infrequent to dominate the evolution of the observable dust mass.

\subsection{Delivering planetesimals to inner resonant planets} \label{sec:inner}

The radial broadening process studied here affects the planetary system more generally, as it advances the reach of planetesimals to regions beyond their initial site of formation. While radial broadening from eccentricity excitation alone is limited, incorporating the broadening in the semimajor axis distribution approximately doubles the overall belt width. 
For sufficiently massive disks, it is possible they will widen enough to encounter planets initially far removed from the ring. 

Among the notable findings of exoplanet demographics from transiting surveys such as Kepler is the observation that the majority of pairs of transiting planets found in $<$100\,Myr-old systems are in orbital resonance ($\sim$70\% in resonance), whereas when found in $>$1\,Gyr-old systems these inner planets are typically out of resonance ($\sim$15\% in resonance, \citealp{Dai2024}). The prevalence of resonance chains at formation can occur from resonance locking during planet migration in the protoplanetary disk \citep{Terquem2007}, in which case the subsequent termination of these resonance requires an explanation. 

Recent theories have argued that planetesimals with a few percent of the planets' mass are sufficient to repel pairs of planets and disrupt commensurability \citep{Hadden2026, Choksi2026, LoRusso2026, Goldberg2026}, although it is not entirely unclear at this stage where these planetesimals come from. It has also been suggested that inner resonance disruption could be achieved by outer companions that secularly perturb the planets \citep{Ogihara2026}, or companions that perturb a distant outer planetesimal reservoir and send a sufficiently large planetesimal mass flux towards the inner planets ($m \sqrt{N} \sim 50 \, M_\oplus$, \citealp{Li2026}). 

Here we consider the presence of a planetesimal ring in such systems, whose radial spreading could naturally deliver planetesimals much interior to the initial ring location even in the absence of external perturbers. Debris disk radial structures have been tentatively suggested to evolve on similar $\sim$100\,Myr timescales \citep{Han2025}, and this work further suggests such an evolution to be theoretically plausible and intrinsic to the disk. Speculatively, as the dynamical timescales of the inner and outer planetary systems differ by 3 orders of magnitude, the coincident resonant breakup and debris disk radial evolution timescales could suggest at some high level that their evolution may be somehow linked. 

\begin{figure}
    \centering
    \includegraphics[width=1.0\linewidth]{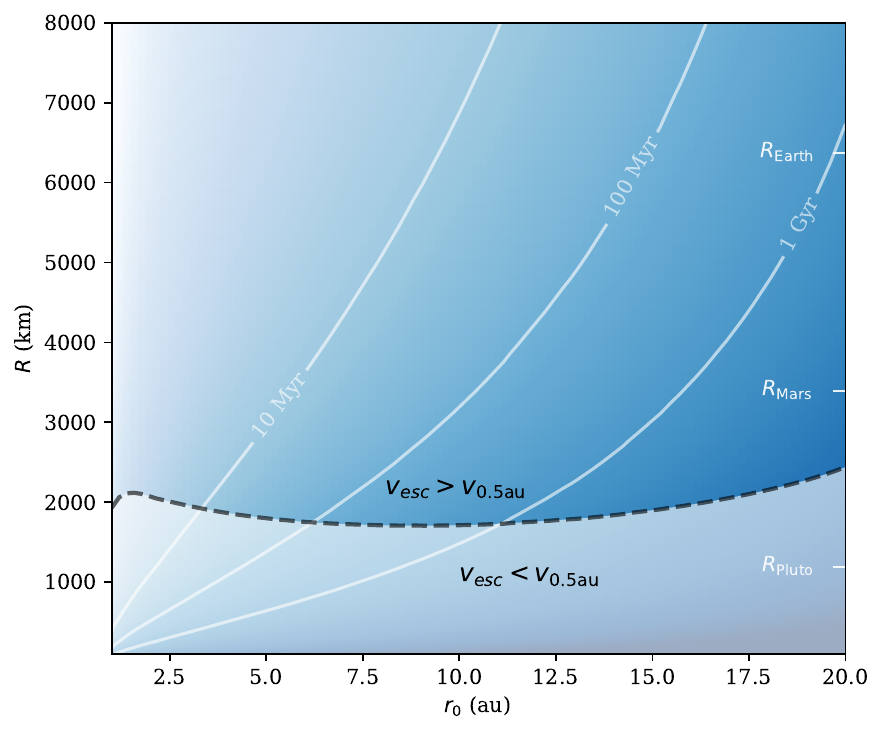}
    \caption{The timescale required for a 10\,$M_\oplus$ ring of planetesimals around a 1\,$M_\odot$ star to radially diffuse 0.1\,$M_\oplus$ into the inner 0.5\,au. The planetesimal density is assumed to be 4\,g\,cm$^{-3}$. The black dashed line shows the boundary where the inner edge is just reachable by the radially diffusing disk, below which the escape velocity is less than the critical escape velocity ($v_\mathrm{0.5au}$) and planetesimals are not deliverable at this level. }
    \label{fig:reach_1au}
\end{figure}

Observationally, about half of resolved debris rings have a fractional radial FWHM above 0.5 at present (Fig.~\ref{fig:histogram}), and about 20\% are inferred to reach or have reached a fractional width of above 1 at an age of 100\,Myr. If the extended inner tail of a debris disk, corresponding to some of the most highly scattered planetesimals in the disk, were to advance as far inwards as to touch the terrestrial region, then planetesimals initially formed in stable regions far from the planets could still supply scattering bodies that contribute to resonance disruption at ages of several 100\,Myr. 

To evaluate the feasibility of the outer planetesimal delivery scenario, we calculated the timescale over which the radial profile extends a Mars mass of planetesimals below $r_e = 0.5$\,au by numerically integrating the radial profile given by Eq.~\eqref{eq:aekernel}, assuming Gaussian $a$ and Rayleigh $e$ in equipartition. 
For a 10\,$M_\oplus$ disk, this corresponds to its radially innermost 1\% of mass, comparable to the $\sim$few percent planet mass thought to be required for resonance disruption \citep{Hadden2026, Choksi2026, LoRusso2026}. 

Assuming the radial evolution described by Eq.~\eqref{eq:rn}, the timescale required to excite $\tilde{r}_{\sigma} = \Delta r_{\sigma}/r$ is
\begin{equation}
    \label{eq:tin}
    \begin{split}
        t_\mathrm{wid} = {} & 2 \, \mathrm{Myr}
        \left( \frac{\tilde{r}_{\sigma}}{0.1} \right)^{5}
        \left( \frac{M_\mathrm{disk}}{1 \ M_\oplus} \right)^{-1} \\
        & \times \left( \frac{m}{1 \ M_\mathrm{Pluto}} \right)^{-1}
        \left( \frac{M_*}{1 \ M_\odot} \right)^{3/2}
        \left( \frac{r_0}{1 \ \mathrm{au}} \right)^{3/2},
    \end{split}
\end{equation}
where $r_0$ is the central radius of the belt. 

Additionally, for $r_e$ to be reachable in the first place, the velocity dispersion required to reach this radial width must satisfy $v_\mathrm{swm} < v_\mathrm{esc}$, giving rise to the condition
\begin{equation}
    \label{eq:cin}
    \begin{split}
        R_\mathrm{wid}
        \gtrsim {} &
        1500 \ \mathrm{km} \left( \frac{\tilde{r}_{\sigma}}{0.1} \right)^{1}
        \left( \frac{\rho}{4 \ \mathrm{g}\,\mathrm{cm}^{-3}} \right)^{-1/2}\\
        & \times \left( \frac{M_*}{1 \ M_\odot} \right)^{1/2}
        \left( \frac{r_0}{1 \ \mathrm{au}} \right)^{-1/2}.
    \end{split}
\end{equation}

We plotted the relations in Eq.~\eqref{eq:tin} and \eqref{eq:cin} within the $r_0$--$R$ plane in Fig.~\ref{fig:reach_1au} for a 10\,$M_\oplus$ disk orbiting a solar-mass star, assuming a planetesimal density of 4.0\,g\,cm$^{-3}$, which is approximately the density of Mars. 
Within $\sim$10\,au, the minimum planetesimal size tends to decrease with radius as it takes a smaller velocity perturbation to induce the same eccentricity for a body on a further out orbit than it does for a closer in body, however at large radii this is counteracted by the larger fractional width required of the ring, causing the threshold planetesimal mass to instead increase. The timescale required to reach the effective $\tilde{r}_{\sigma}$ increases towards large radii due to both lower stirring rates further out and larger velocity dispersions required for the radial profile to extend towards the inner system. 

Accounting for both the minimum planetesimal mass and the stirring timescales, we find that a 10\,$M_\oplus$ belt of Mercury-sized bodies at a few au can plausibly spread far enough to deliver a few such bodies (0.1\,$M_\oplus$) within $\sim$10\,Myr. The same total belt mass but with larger constituent bodies will achieve this faster in proportion to the mass of each body. Once inside a fraction of an au, the collisional timescale with any terrestrial planets is rapid and on the order of 0.01 to 0.1\,Myr. From this simple scaling analysis, such planetesimals formed in a massive asteroid-belt analogue exterior to the terrestrial planets could plausibly contribute to reshaping the orbital configuration of inner planets. If such is the case, a fraction of these viscously spreading planetesimals would inevitably cross orbits with any terrestrial planets, possibly leading to giant impacts that manifest as young, inner extreme debris disks \citep{Schneiderman2021, Su2022, Su2026}. 

Alternatively, it is possible that a more massive and further out Kuiper-belt analogue could also deliver a similar planetesimal flux over the same timescale. A belt of 1000 Mars-sized bodies at 40\,au, which would total 100\,$M_\oplus$, would be able to deliver 0.1\% of its mass into the inner 0.5\,au over 10\,Myr. Of the resolved debris disk sample studied in Sec.~\ref{sec:observations}, 40\% are inferred to have $m M_\mathrm{disk}$ above this level. 
The delivery rate could be further enhanced if the outer belt were to be perturbed by a companion, such as a giant planet or stellar companion \citep{Li2026}, or any embedded super-Earth stirrers as suggested by the disks with the highest $m M_\mathrm{disk}$ of $10^3 \, M_\oplus^2$(Sec.~\ref{sec:mass}). Likewise, if planetesimals were to encounter planets on their way in, these planets could accelerate delivery of planetesimals inwards to rates beyond those calculated above. The Solar System serves as another example in which there exist planetesimals with pericentres close to the Sun despite the low mass of the Kuiper belt \citep{Levison1997}.

While an asteroid-belt analogue at a few au is unresolvable with ALMA, one naturally wonders in the Kuiper-belt analogue scenario whether there is observational evidence of planetesimal belts whose inner edge tail extends close to the star. In extrasolar systems, the radial profile gallery fitted to high-resolution debris disk observations (Fig.~3 in \citealt{Han2026}) suggests disks with extended inner edges are common, and even nominally narrow disks often exhibit extended inner tails that reach close to the star. The disk masses assumed here may appear to be large, and almost certainly more massive than the terrestrial planets of a typical system combined, but it is well within reason from the perspective of available solids during planet formation \citep{Wyatt2002, Krivov2021}, scale height-inferred masses (\citealp{Zawadzki2026, Chiang2026}; Jankovic et al. submitted), and estimates of the size and number of bodies for observable giant impact signatures in $\beta$~Pic to be likely \citep{Han2023}. It is thus possible that these well-resolved debris disks represent the kind of radial profiles that are able to extend their reach to the terrestrial region to levels thought to induce resonance breakup. 

But do the known transiting systems host debris disks? The debris disk detection rates around (candidate) transiting planets appears to be as low as among the general stellar population \citep{Kennedy2012}. Debris disks, even if initially massive, are generally difficult to detect around old stars, which are common in exoplanet transit surveys. Indeed, the bulk of the ALMA-resolved debris disk sample represents the biggest and brightest disks, and reaching the the sensitivity to detected 2$\sigma$ radial profile tails in more ordinary disks remains challenging. The ``opposite'' effect is also possible, in which rather than the mass in any collisional cascade having been depleted at Gyr timescales, much of the mass is simply locked in the largest bodies that rarely undergo collisions due to their low number density. It could be that through oligarchic growth and any subsequent giant impacts during the final phases of planetesimal formation \citep{Kokubo1998}, only a limited amount of debris is left in a steady-state collisional cascade for some of these massive belts to be visible, and they do not manifest as the bright debris disks targeted by ALMA until fresh debris is injected through sporadic giant impacts. 

Regardless of whether these planetesimals originate from asteroid- or Kuiper-belt analogues, if it is true that a large fraction of resonant planet chains are disrupted by viscously spreading planetesimals, it is expected that outer planets within a given chain preferentially break from resonances earlier than inner ones do, which future studies may wish to test. Future work could also investigate this scenario by targeting systems that simultaneously host planets and outer debris disks, searching for or rejecting any correlations between them.

\subsection{Ejecting free-floating planets} \label{sec:ejection}

\begin{figure}
    \centering
    \includegraphics[width=1.0\linewidth]{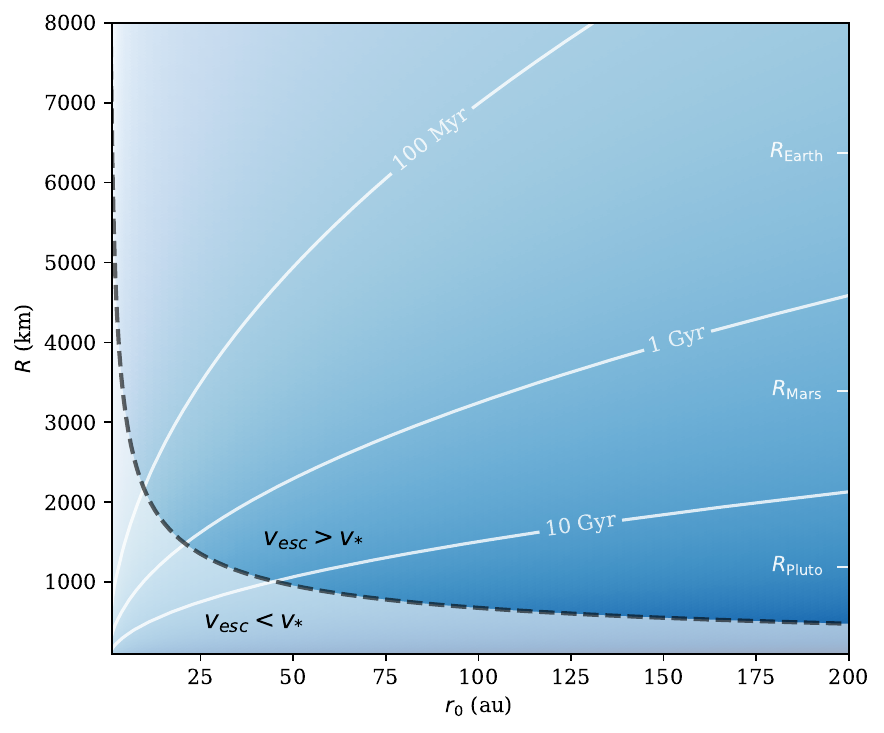}
    \caption{The timescale required for a ring of massive planetesimals undergoing mutual scattering to eject 1\% of its bodies into interstellar space. The total disk mass is assumed to be 100\,$M_\oplus$, the stellar mass 1\,$M_\odot$, and the planetesimal density 4\,g\,cm$^{-3}$. The black dashed line shows the boundary below which the escape velocity is less than the critical escape velocity ($v_*$) and bodies are not ejected at 1\% levels. }
    \label{fig:escape_star}
\end{figure}

The inner and outer edges of planetesimal belts diffuse simultaneously. For sufficiently massive disks, it is possible for a fraction its planetesimals to be ejected. While recent encounters with interstellar objects in the Solar System such as Oumuamua \citep{Meech2017} have hinted at the presence of a substantial population of free-floating km-sized bodies \citep{Do2018}, microlensing surveys have been detecting a population of ``free-floating planets'', where terrestrial-sized planets are not accompanied by detections of host stars typically within $\sim$5--10\,au \citep{Mroz2020, Koshimoto2023}. 

Motivated by the suggestion of free-floating terrestrial planet-sized bodies, we investigate the possibility for disks of massive planetesimals to eject a small fraction of its constituent bodies via mutual scattering. We consider a disk with mass dominated by bodies similar in size to the terrestrial planets such as Mercury or Mars. While this is large for a ``planetesimal'', they could plausibly exist in debris disks with masses of 10s to $\sim$100 $M_\oplus$ (\citealp{Zawadzki2026, Chiang2026}, Jankovic et al. submitted), and have been invoked to explain long-lasting asymmetries possibly caused by giant impacts within cold outer belts \citep{Telesco2005, Jackson2014, Han2023}.

Two conditions must be met for such a planetesimal to be ejected from the disk. Firstly, the planetesimal's escape velocity must be large enough such that sufficiently high random velocities can be induced for some fraction $\psi$ of the planetesimals to leave the system. A planetesimal is ejected when its total velocity satisfies $v_\mathrm{tot} = ( v_\mathrm{K}^2 + v^2 - 2 v_\mathrm{K} v \cos{\alpha} )^{1/2} > v_\mathrm{esc}^* = \sqrt{2} v_\mathrm{K}$, where $v$ is the random velocity relative to local circular Keplerian, $\alpha$ is the angle between $v$ and $v_\mathrm{K}$ and $v_\mathrm{esc}^*$ is the escape velocity of the star. For a randomly oriented $v$ vector in space with a Rayleigh-distributed magnitude, the mean random velocity $v_\mu$ must satisfy $v_\mu / v_\mathrm{K} > \phi$ to eject a fraction $\psi$ of the planetesimals, where $\phi = 0.24$ for $\psi = 1\%$ from numerical simulations. Since the maximum $v_\mu$ reached is approximately $v_\mathrm{esc}/\sqrt{2}$, this condition is equivalent to $v_\mathrm{esc} / v_\mathrm{K} \gtrsim \sqrt{2} \, \phi$, or 
\begin{equation}
    \label{eq:cout}
    \begin{split}
        R_\mathrm{ej}
        \gtrsim {} &
        700 \ \mathrm{km} \left( \frac{\phi}{0.24} \right)^{1}
        \left( \frac{\rho}{4 \ \mathrm{g}\,\mathrm{cm}^{-3}} \right)^{-1/2}\\
        & \times \left( \frac{M_*}{1 \ M_\odot} \right)^{1/2}
        \left( \frac{r_0}{100 \ \mathrm{au}} \right)^{-1/2}.
    \end{split}
\end{equation}

The second condition is that the stirring timescale should be shorter than the age of the system. Using Eq.~\eqref{eq:vn}, the timescale for $v_\mu / v_\mathrm{K}$ to reach $\phi$ is
\begin{equation}
    \label{eq:tout}
    \begin{split}
        t_\mathrm{ej} = {} & 100 \, \mathrm{Myr}
        \left( \frac{\phi}{0.24} \right)^{5}
        \left( \frac{M_\mathrm{disk}}{100 \ M_\oplus} \right)^{-1} \\
        & \times \left( \frac{m}{1 \ M_\oplus} \right)^{-1}
        \left( \frac{M_*}{1 \ M_\odot} \right)^{3/2}
        \left( \frac{r_0}{100 \ \mathrm{au}} \right)^{3/2}.
    \end{split}
\end{equation}

We plotted the two relations in Fig.~\ref{fig:escape_star} for a 100\,$M_\oplus$ disk around a solar-mass star. We find that for a belt to eject 1\% of its mass on $\sim$Gyr timescales, the belt is required to be located at 10s to $\sim$100\,au, with the most massive bodies being at least the size of the Moon to Mars, depending on the belt location. 
A belt of even more massive bodies could achieve this on $\sim$100\,Myr timescales. 
This is satisfied by close to half of resolved debris disks (Sec.~\ref{sec:mass}). These ejection fluxes could be further increased by any embedded super-Earth stirrers, or if there were to be planets external to the disk that further perturb the planetesimals. Overall, this analysis illustrates that based on simple scaling relations of a self-stirred belt, it is plausible for a massive disk with reasonable if not common physical properties to plausibly eject several terrestrial planet-sized bodies during its long-term evolution. 

If such a picture is true, this would imply that even in the absence of perturbing planets, debris disks could themselves eject planet-sized bodies and contribute to what appears to be a free-floating planet population. This would loosen requirements on the planetary system architecture or the presence of giant planets required for inner terrestrial planets to be scattered \citep{Barclay2017, Coleman2025}, making the ejection of terrestrial planet-sized bodies a general feature for any massive disk-hosting planetary system, as long as Mercury- or Mars-sized bodies dominate the mass of massive debris disks. This would also imply that a large fraction of free-floating planets could have formed beyond the snow line, thereby being more icy in composition. 

If some fraction of debris disks truly host bodies as massive as Mars, then despite the possibility of ejection explored above, the vast majority of these bodies are expected to remain within the disk, or at least be loosely bound. A large fraction of planets detected via microlensing without a host star would then likely represent large planetesimals in the outer planetary system, rather than truly unbound bodies. This is similar to the case of any wide-orbit and ejected planets from planet-planet scattering, for which \citet{Hadden2026a} suggested that approximately half of microlensing planets without host stars are bound. While bulk densities and thus composition may be difficult to measure for free-floating planets, statistics on their size distribution from upcoming microlensing surveys, such as with the \textit{Roman Space Telescope}, may be able to tell whether the free-floating planet population is consistent with that expected from debris disk bodies.

\subsection{Limitations and future work} \label{sec:limitations}
There are several limitations and caveats worth re-emphasising. Firstly, several effects are neglected. We have ignored the effect of collisions between the most massive bodies in the belt and any velocity damping this could have caused \citep{Lithwick2007}, which is mainly motivated by the low collision rates between them, as discussed in Sec.~\ref{sec:collisions}. Our simulations are relatively well-fitted by analytical expectations that ignore such collisions. The disk's own gravitational potential is also ignored, as it is not thought to play an important role in self-stirring \citep{Pearce2025}. 

We have also ignored any effect due to gas. In interpreting the width of protoplanetary rings observed as a reflection of the planetesimal distribution, we have assumed that these planetesimals are too large to be affected by aerodynamic gas drag in the primordial gas disk, but too small to be influenced by back reaction from any spiral waves that they induce on the gas disk. This interpretation also assumes that the dust rings are a reflection of stirring by these large planetesimals, such as due to a significant fraction of dust originating from a collisional cascade stirred by the most massive planetesimals already formed. 
The presence of planets would naturally affect scattering rates. For example, \citet{Levison2011} found that self-stirred planetesimal disks could over time couple to multi-resonant planets, inducing delayed instabilities over timescales similar to the late heavy bombardment.

Secondly, we have assumed that the millimetre-sized pebbles in debris disks observed by ALMA are generated from the collisions of smaller planetesimals that are abundant enough to collide in steady state but which do not dominate the mass, and that they trace the distribution of the most massive bodies in the disk that drive the belt's radial broadening and vertical thickening. Under this assumption, we modelled only a single planetesimal population, ignoring any dynamical friction from smaller bodies. In reality, bodies from dust to planetesimals are expected to exist in a continuous (though not necessarily smooth) size distribution. Future work may wish to incorporate a two-population model that focuses on the evolution of smaller planetesimals in steady-state collisions under the influence of massive invisible planetesimals, and the effect of sporadic collisions between large planetesimals even if they were to occur at low frequencies. 

Thirdly, we have not modelled any variations in the scaling relations as the disk becomes very broad ($\Delta r / r \sim 1$). As the disk broadens, the inner regions may experience faster stirring rates and the outer regions slower ones, resulting in an asymmetric radial profile that that departs from Gaussian. Indeed, the radial profiles of debris rings show diversity in shape \citep{Han2026}, which is particularly the case among the broadest disks, noting also that this could be in part due to those disks being easier to observationally resolve. Future work may wish to characterise this intrinsic asymmetry and model in detail its effect on the rate of planetesimal delivery to the terrestrial region. 

Finally, while we have considered planetesimal formation in narrow rings, such as in dust traps in protoplanetary disks, some models have suggested that stationary pressure bumps favour the formation of planets rather than planetesimal belts \citep{TLau2022}. It has been suggested that dust trap migration may be necessary to prevent such an outcome, in which case a radially broader planetesimal belt is expected to form, unless if perturbed by other forces, such as resonances with already-formed planets \citep{TLau2025}. Previous theoretical studies have proposed a variety of effects that could cause the site of planetesimal formation to migrate, such as the receding photoevaporation front of the primordial gas disk \citep{TLau2025, LiChiang2026}, perturbations from already-formed planets \citep{Miller2021} or a traffic jam in pebble drift \citep{Jiang2023}. However, if radially broad formation channels are indeed prevalent, additional explanations may be required for why the relative narrow radial width of debris rings relative to their scale height appears to leave little room for a wide initial semimajor axis distribution, as noted in Sec.~\ref{sec:h_evolution}.

It is possible that radially narrow and broad planetesimal formation channels both operate in a given planetary system. After all, the observed radial profiles of debris rings are rarely perfectly Gaussian \citep{Han2026}. Speculatively, the prevalence of narrow debris rings that sit atop radially extended low-amplitude emission could represent the compounded radial profile of planetesimals formed via radially narrow and broad channels. These faint, broad components are not reflected in the fractional width measurements considered in Sec.~\ref{sec:observations} due to their contribution to the radial profile being low in amplitude. 

With a robust understanding of the intrinsic radial evolution enabled by this and future work suggested, further studies can embark on including the additional complexity of perturbing planets, such as how planets carve into disks as they spread towards the planet, any migration that this may induce on the planet, and how such planets could alter the trajectory of planetesimals on their path towards the inner system.

\section{Conclusions} \label{sec:conclusions}
We investigated how narrow planetesimal belts formed in protoplanetary disks evolve in the absence of perturbing planets. Unlike the case of small planetesimals stirred by a massive population, we focused on the orbital evolution of the massive population themselves with analytic scaling and N-body simulations. The main findings are summarised below. 

\begin{enumerate}
    \item A belt of mutually gravitating planetesimals in an initially narrow ring spreads out radially and vertically over time. Unlike small planetesimals stirred by orbitally stationary massive bodies, the dispersion of the orbital elements of the massive bodies themselves scales as $t^{1/5}$, with $\Delta a_\sigma/a_0 : e_\mu : i_\mu \approx 3 : 2 : 1$. The width of the resulting radial and vertical profiles likewise scales as $t^{1/5}$, with $\Delta r_\mathrm{FWHM} / r = 10 \, h_\sigma$ at equipartition. The semimajor axis distribution is maintained as Gaussian, whereas the eccentricity and inclination are Rayleigh distributed. The resulting radial and vertical profiles are approximately Gaussian. 
    
    \item We examined the population of resolved rings in protoplanetary and debris disks, motivated by a hypothesised picture with narrow protoplanetary rings as the starting point for planetesimals that go on to populate debris disks, rather than a wide power-law initial debris disk profile, with gaps and edges subsequently sculpted by planets. While debris rings as a population have a larger $\Delta r / r$ than protoplanetary rings, scaling debris rings back to a few Myr closely matches the observed $\Delta r / r$ distribution of protoplanetary rings. The debris ring population lies within a factor of a few from the equipartition relation between the radial width and scale height, consistent with planetesimal formation occurring over a semimajor axis range as narrow as the (likely low) eccentricity and inclination dispersions. 
    
    \item Assuming ALMA-resolved debris rings formed narrow, the total planetesimal mass times the mass of each individual planetesimal, $mM_\mathrm{disk}$, required to broaden them to their widths observed at present exhibits a stellar mass dependence, peaking at approximately 1.5--2\,$M_\odot$, in synchrony with the giant planet occurrence rate as a function of stellar mass. Even when assuming the lower bound of large-body masses based on these $mM_\mathrm{disk}$ measurements, effectively no resolved debris ring has reached their equilibrium radial width or vertical height. The large scatter in $mM_\mathrm{disk}$ suggests diverse planet formation outcomes at tens of au and beyond, with the largest $mM_\mathrm{disk}$ rings, typically found around A and early F stars, appearing to require formation of either a dozen embedded super-Earths or external perturbing planets. 

    \item The radial broadening of debris disks is a plausible mechanism of planetesimal delivery to the terrestrial region, which could drive the breakup of resonance chains of planets observed by the Kepler mission \citep{Dai2024}. A 10\,$M_\oplus$ asteroid-belt analogue consisting of Mercury-sized bodies initially at a few au, or a 100\,$M_\oplus$ Kuiper-belt analogue consisting of Mars-sized bodies at tens of au, can deliver $\sim$0.1\,$M_\oplus$ to the terrestrial planet region within $\sim$10\,Myr, suggesting that even in the absence of any left-over planetesimals that survive among a resonant chain of planets, exterior debris disks could still supply planetesimals to trigger instabilities in the inner system. 

    \item Mutual scattering between massive bodies in a debris disk can eject a fraction of its bodies that contribute to the ``free-floating planet'' population. Approximately $\sim$1\% of an 100\,$M_\oplus$ disk of Mars-sized bodies at tens of au is ejected on $\sim$Gyr timescales, suggesting that each massive debris disk could eventually contribute at least a few terrestrial planet-sized bodies that could be picked up by microlensing surveys. Such an ejection mechanism does not require planetary or stellar companions, and is applicable as long as Moon-sized bodies or above dominate the mass of massive debris disks. The size distribution derived from microlensing surveys may be able to inform whether massive debris disks are a dominant source of free-floating or wide-orbit planets. 
    
\end{enumerate}


\begin{acknowledgments}
YH is grateful for discussions with Eugene Chiang and Daniel Tamayo at the International Conference on Exoplanets and Planet Formation in Shanghai on the plausibility of analytical and numerical approaches. YH is also grateful for the opportunity to collaborate with Sebastian Marino and the ARKS team on an interesting comparison between protoplanetary and debris disks that partly inspired to this work. We acknowledge the use of GitHub Copilot and Google Gemini for coding assistance. YH gratefully acknowledges support by a Caltech Barr Fellowship. 
\end{acknowledgments}

\begin{contribution}
YH considered scaling relations, performed simulations and wrote the manuscript. All authors contributed to scientific interpretation and manuscript editing. 

\end{contribution}


\software{\texttt{NumPy} \citep{numpy}, 
          \texttt{SciPy} \citep{scipy},
          \texttt{Matplotlib} \citep{matplotlib},
          \texttt{Astropy} \citep{astropy:2022},
          \texttt{GENGA} \citep{Grimm2014, Grimm2022}
          }


\appendix
\section{Establishing equipartition} \label{sec:equipartition}
\begin{figure}
    \centering
    \includegraphics[width=0.8\linewidth]{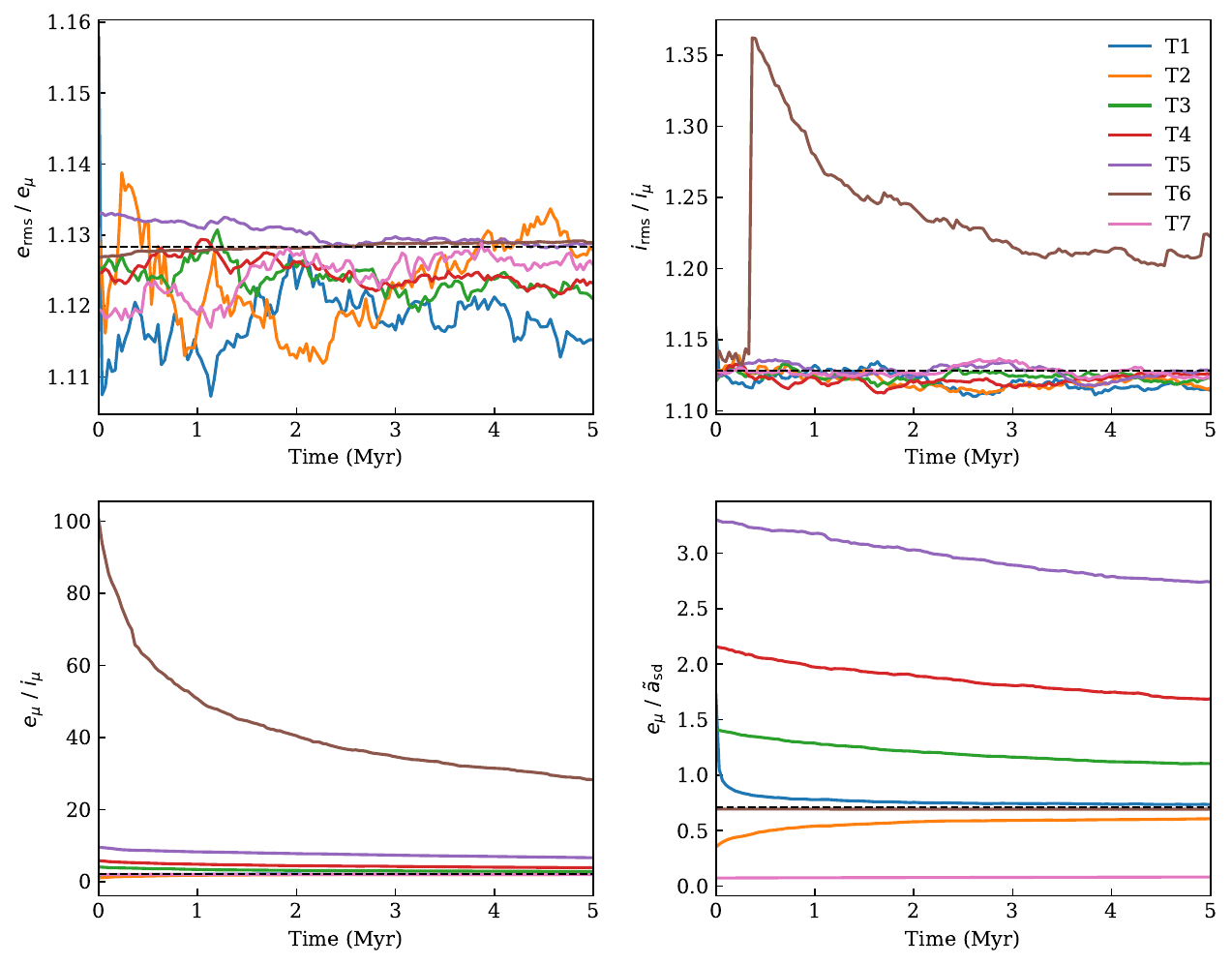}
    \caption{Evolution of $\tilde{a}$, $e$ and $i$ ratios used to gauge the ``Rayleigh-ness'' of $e$ and $i$ (top panels) and the rate at which the three orbital elements approach equipartition (bottom panels). The dashed lines represent the theoretical equipartition ratios. }
    \label{fig:out_of_equipartition}
\end{figure}

As this study focuses on the case of equipartition between semimajor axis, eccentricity and inclination, we have not addressed in detail whether the initial conditions of planetesimals formed are in equipartition, and if not, the timescale over which they establish equipartition. A comprehensive investigation of this topic is beyond the scope of this study, however we ran a simple suite of N-body simulations with out-of-equipartition initial conditions to gauge the cause of their evolution on some simplified level. These simulations are summarised in Table~\ref{tab:nbody2} and plotted in Fig.~\ref{fig:out_of_equipartition}. 

These trial simulations mainly suggest three conclusions. Firstly, a Rayleigh distribution of $e$ and $i$ and a Gaussian distribution of $a$ are quickly established for our disk setup even if their initial distributions are uniform (T1 in Fig.~\ref{fig:out_of_equipartition} top panels). 

Secondly, the Rayleigh distribution is maintained even if $e$ and $i$ are moderately out of equipartition (within a factor of a few). However, we tested a heavily out-of-equipartition scenario with $e \gg i$, motivated by the suggestion that such configurations could exist in debris disks \citep{Chiang2026}. In such a scenario, $i$ is no longer Rayleigh distributed, which is sustained until the end of our simulation at 5\,Myr (T6 in Fig.~\ref{fig:out_of_equipartition} top right panel), similar to what \citet{Chiang2026} found in the case of small planetesimals stirred by massive bodies. However, if the initial dispersion in $a$ is significantly larger than that of $e$ and $i$ (T7 in Fig.~\ref{fig:out_of_equipartition}), then $e$ and $i$ still remain Rayleigh distributed. 

Thirdly, the system tends towards equipartition more slowly if $i$ starts off suppressed relatively to equipartition (T1 and T2 in Fig.~\ref{fig:out_of_equipartition} bottom panels), compared to if $i$ is initially enhanced (T3--T6). 
If the dispersion in $\tilde{a}$ is initially enhanced relative to equipartition with $e$ and $i$, the rate at which $e$ and $i$ catch up with $\tilde{a}$ is slow, increasing by a negligible amount relative to the order-of-magnitude larger $\tilde{a}$ over 5\ Myr in our simulation T7. 
While this work has focused on the case of equipartition, future work may wish to investigate out-of-equipartition evolution in more detail than what is explored here. 

\begin{table}[htbp]
\centering
\caption{Initial parameters for N-body simulations out of equipartition.}
\label{tab:nbody2}
\begin{tabular}{cccc}
\toprule
Run & $\tilde{a}_0$ & $e_0$ & $i_0$ \\
\midrule
T1 & Uniform  0\,--\,0.01  & Uniform  0\,--\,0.01 & Uniform  0\,--\,0.01 \\
T2 & Gaussian $\sigma = 0.01 \sqrt{\pi}$  & Rayleigh $\sigma = 0.005$ & Rayleigh $\sigma = 0.005$ \\
\rowcolor{gray!20}[\tabcolsep]
i & Gaussian $\sigma = 0.01 \sqrt{\pi}$   & Rayleigh $\sigma = 0.01$  & Rayleigh $\sigma = 0.005$   \\ 
T3 & Gaussian $\sigma = 0.01 \sqrt{\pi}$   & Rayleigh $\sigma = 0.02$   & Rayleigh $\sigma = 0.005$  \\
T4 & Gaussian $\sigma = 0.01 \sqrt{\pi}$   & Rayleigh $\sigma = 0.03$   & Rayleigh $\sigma = 0.005$  \\
T5 & Gaussian $\sigma = 0.01 \sqrt{\pi}$   & Rayleigh $\sigma = 0.05$   & Rayleigh $\sigma = 0.005$  \\
T6 & Gaussian $\sigma = 0.1 \sqrt{\pi}$    & Rayleigh $\sigma = 0.1$    & Rayleigh $\sigma = 0.001$  \\
T7 & Gaussian $\sigma = 0.1 \sqrt{\pi}$    & Rayleigh $\sigma = 0.01$    & Rayleigh $\sigma = 0.005$  \\
\bottomrule
\end{tabular}
\tablecomments{Parameters not listed here are identical to those in Table~\ref{tab:nbody} Run i. A subset of parameters from Run i are displayed here for comparison. 
}
\end{table}

\bibliography{references}
\bibliographystyle{aasjournalv7}



\end{document}